\documentclass[a4paper, 11pt]{article}
\pdfoutput=1
\usepackage{jheppub}
\makeatletter
\def\@fpheader{}
\makeatother

\usepackage{graphicx,wrapfig,float,slashed}
\usepackage{bm}
\usepackage{amsmath,amssymb,dsfont,epsfig,graphicx}
\usepackage{mathtools} 
\usepackage{multirow}
\usepackage{blkarray}
\usepackage{epstopdf}
\usepackage{ragged2e}
\usepackage[utf8]{inputenc} 
\usepackage{cancel}
\usepackage[usenames,dvipsnames,table]{xcolor}
\usepackage[toc]{appendix}
\usepackage[normalem]{ulem}
\usepackage{enumitem}
\usepackage[font=small,skip=1pt]{caption}
\allowdisplaybreaks
\usepackage{pifont}
\usepackage{subcaption}
\usepackage{footnote}

\newcommand{\beq}{\begin{eqnarray}}
\newcommand{\eeq}{\end{eqnarray}}

\begin{document}

\title{Extended Color Twin Higgs}

\author[1]{Brian Batell,}
\emailAdd{batell@pitt.edu}
\author[2]{Thomas Cochran,}
\emailAdd{tcochra1@byu.edu}
\author[2,3]{Logan Page,}
\emailAdd{lop@uoregon.edu}
\author[2]{and Christopher B. Verhaaren}
\emailAdd{verhaaren@physics.byu.edu}
\affiliation[1]{Pittsburgh Particle Physics Astrophysics and Cosmology Center, 
Department of Physics and Astronomy, University of Pittsburgh, Pittsburgh, PA 15260, USA}
\affiliation[2]{Department of Physics and Astronomy, Brigham Young University, Provo, UT, 84602, USA}
\affiliation[3]{Department of Physics, University of Oregon, Eugene, OR, 97403, USA}

\abstract{
 We describe a novel variation of the mirror Twin Higgs model in which the color gauge group in both sectors is extended to SU(4$)_c$ and spontaneously broken to SU(3$)_c$ exclusively in the visible sector. Through this process, the mirror $Z_2$ symmetry is spontaneously broken, allowing for a phenomenologically viable electroweak vacuum alignment. 
 This structure produces interesting collider signatures, including heavy vectors and fermions with fractional electric charges. The twin sector, with unbroken SU(4$)_c$, produces interesting cosmological characteristics, such as the possibility to reduce $\Delta N_\text{eff}$ and stable spin-0 baryons. The enlarged top quark sector required by the extended color gauge symmetry preserves naturalness, with even less tuning than the original twin Higgs in many circumstances.
}

\preprint{}
\arxivnumber{}

\flushbottom
\maketitle
\flushbottom

\section{Introduction\label{sec:Intro}}
Among the major insights from the Large Hadron Collider (LHC) are the presence of a Higgs boson consistent with Standard Model (SM) expectations and the apparent absence of new TeV-scale states, challenging conventional solutions to the hierarchy problem such as supersymmetry and Higgs compositeness. In light of this situation, the paradigm of neutral naturalness has emerged as an appealing alternative for stabilizing the electroweak scale; see Ref.~\cite{Batell:2022pzc} for a review. The most studied realization of these ideas is the Twin Higgs~\cite{Chacko:2005pe} framework, which protects the Higgs mass from large corrections up to UV scales of order 10 TeV by positing a hidden copy of the SM related by a $Z_2$ mirror symmetry. Along with an approximate global SU(4) symmetry of the Higgs potential, the little hierarchy problem can be addressed with only a mild tuning at the 10$\%$ level.

To generate a phenomenologically viable vacuum with a hierarchy between the SU(4) breaking scale and the electroweak scale, the $Z_2$ symmetry must be broken. Such $Z_2$ breaking can be introduced in various ad hoc ways, including softly via symmetry breaking mass terms in the Higgs potential~\cite{Chacko:2005pe}, or in a hard manner by eliminating certain twin-sector states~\cite{Craig:2015pha} or through a mismatch between SM and twin-sector couplings~\cite{Barbieri:2016zxn}. Alternatively, one can study the possible dynamics responsible for breaking $Z_2$, see Refs.~\cite{Beauchesne:2015lva,Harnik:2016koz,Yu:2016bku,Yu:2016swa,Jung:2019fsp,Bittar:2024ryj} for some representative studies. In this vein, Refs.~\cite{Batell:2019ptb,Liu:2019ixm,Batell:2020qad} considered the joint spontaneous breaking of twin sector gauge symmetries and $Z_2$. These models preserve Higgs naturalness while opening up additional interesting possibilities for precision measurements and collider searches, as well as twin sector cosmology and dark matter~\cite{Curtin:2021spx,Kilic:2021zqu}. In this work we consider the related idea of spontaneously breaking a gauge symmetry in the visible sector, while leaving it unbroken in the twin sector. As in \cite{Batell:2019ptb,Liu:2019ixm,Batell:2020qad}, this necessarily brings about the dynamical breaking of $Z_2$.

The possibility that the SU(3$)_c$ gauge symmetry of the SM arises from a larger color group, such as SU(4$)_c$, that is spontaneously broken at some high scale was considered long ago~\cite{Foot:1989bu,Foot:1989zk,Foot:1990kr,Pleitez:1993vf,Foot:2011xu}. We implement this structure in a twin Higgs context, but leave the SU(4$)_c$ twin color group unbroken. The symmetry breaking in the visible sector gives rise to beyond the Standard Model (BSM) particles that can be discovered at current or future colliders. These include new fermions associated with the SM quarks. After color breaking, the fermions in the fundamental representation of the larger color group produce the SM quark fields along with new states whose masses are tied to the color-breaking scale.

These BSM fermions are color neutral and have electric charge of $\pm 1/2$. The LHC searches~\cite{CMS:2012xi,CMS:2024eyx} for fractionally charged particles are still developing, making the short-term discovery prospects for these states exciting. Current sensitivities are only at the few hundreds of GeV scale, but may exceed 1 TeV in the near future. One might worry that TeV scale states with top quark sized coupling to the Higgs negates the natural Higgs mass provided by the twin sector. However, we show that the extended color model remains natural in such cases. Indeed, for much of the motivated parameter space we find that the tuning in our extended color twin Higgs model is typically on par with, and occasionally reduced in comparison to, the standard mirror twin Higgs. This is a consequence of structural features of the model that enforce partial cancellations in the contributions to the Higgs mass parameter, with the enlarged top quark sector required by SU(4)$_c$ playing an essential role. 

That the larger color symmetry persists to low energies in the twin sector also leads to many interesting cosmological consequences. The strong coupling runs more quickly leading to a higher scale of confinement. This can lead to significantly smaller contributions to $\Delta N_\text{eff}$ compared to the standard twin Higgs. The baryons of the twin sector are composed of four quarks. The ground state baryons are bosons with nonzero twin electric charge. These bosons, or combinations of them, could contribute to the cosmological dark matter. 

In the following section we describe the general aspects of the extended color twin Higgs model. Many of the precise details are reserved for Appendices~\ref{app:su4} and~\ref{a.Details} so that Sec.~\ref{sec:model} can focus on the salient structure and results. The experimental constraints and opportunities related to the SU(4$)_c\to$SU(3$)_c$ symmetry breaking are discussed in Sec.~\ref{sec:pheno}. Several of the detailed formulae related to this analysis are recorded in Appendix~\ref{a.Pheno}. In Sec.~\ref{sec:Higgs} we describe the scalar potential of the model and discuss tuning. The details of the tuning calculations for both the mirror twin Higgs and extended color variation are found in Appendix~\ref{ap:tunig}. The cosmological aspects of the extended color model are discussed in Sec.~\ref{sec:cosmo}. This includes an estimate of $\Delta N_\text{eff}$ and a discussion of SU(4$)_c$ baryons. We conclude and offer an outlook on avenues for further work in Sec.~\ref{sec:Con}.

\section{Mirror Twin Higgs with Extended Color\label{sec:model}}
In this section we provide a broad overview of our novel twin Higgs scenario. We focus on modifications from more familiar twin Higgs constructions (the general structure of which is reviewed in Appendix~\ref{ap:tunig}) and an overview of some of the experimental consequences. A more complete description of the physical vectors and fermions are given in Appendix~\ref{a.Details}. The scalar sector of the model is discussed in Sec.~\ref{sec:Higgs}. The visible sector structure is similar to that of~\cite{Foot:1989bu,Foot:1989zk,Foot:1990kr}, but we use different notations for the various BSM fields in order to highlight their connections to SM flavors. 

Our starting point is a mirror twin Higgs-like set up, but one in which the gauge structure in both the visible (or $A$) and twin (or $B$) sectors is SU(4$)_c\times$SU(2$)_L\times$U(1$)_X$. Each sector also includes a new scalar field which can be organized into a doublet
\beq
\Phi=\left( \begin{array}{c}
    \Phi_A \\
     \Phi_B
\end{array}\right)~.
\eeq
Here $\Phi_A$ is in the fundamental representation of SU(4$)_{cA}$ and has U(1$)_{X\,A}$ charge 1/8, while $\Phi_B$ is in the fundamental representation of SU(4$)_{cB}$ with U(1$)_{X\,B}$ charge 1/8. The vacuum of the scalar potential (discussed in more detail in Sec.~\ref{sec:Higgs}) spontaneously breaks the $Z_2$ twin symmetry by giving a VEV to $\Phi_A$, with $\langle\Phi_A\rangle=w$, that breaks SU(4$)_{cA}\times$U(1$)_{X\,A}$ down to SU(3$)_c\times$U(1$)_Y$ but leaves the twin color unbroken, $\langle\Phi_B\rangle=0$. A cross-quartic interaction between the electroweak Higgs and the colored scalars induces an effective $Z_2$-breaking Higgs mass term, enabling a phenomenologically viable alignment of the electroweak vacuum.

The symmetry breaking in the visible sector gives mass to seven of the SU(4$)_{cA}$ gluons. Six of these are grouped into a complex vector $\xi^\mu$ in the fundamental representation of the residual color SU(3$)_c$. As shown in App.~\ref{a.Vecs} this state has a mass near the color-breaking VEV $w$ and an electric charge of 1/6. At colliders, this vector can be pair produced or produced in association with the fourth component of the quark fields.

The remaining massive gluon mixes with the other electrically neutral gauge bosons. It primarily gives rise to a massive $Z'_\mu$, with mass near $w$ and large couplings to quark states. As shown in App.~\ref{a.Details}, the couplings of all neutral vector bosons to a field $\psi$ are given by
\begin{align}
&eQ_\psi A_\mu+Z_\mu\frac{g}{c_W}\left[c_M\left(t^3-s_W^2Q_\psi\right)+\frac{s_Ms_Wc_X}{8s_X}\left(T_c-8\frac{s_X^2}{c_X^2}X_\psi \right) \right]\nonumber\\
&+Z'_\mu\frac{g}{c_W}\left[\frac{c_Ms_Wc_X}{8s_X}\left( T_c-8\frac{s_X^2}{c_X^2}X_\psi \right)- s_M\left(t^3-s_W^2Q_\psi\right) \right]~,\label{eq:neutBosCoup}
\end{align}
where $t^3$ is the third generator of SU(2$)_L$ and $T_c$ is proportional to the fifteenth generator of SU(4$)_c$, see Eq.~\eqref{eq:defTc}. The electric charge is related to the more primal gauge group generators by
\beq
Q_\psi=X_\psi+t^3+\frac18 T_c~.\label{eq:Echarge}
\eeq
Because the $Z'_\mu$ couples to SM quarks it can be produced through Drell-Yan processes at hadron colliders. As shown in Sec.~\ref{sec:pheno} this leads to TeV-scale bounds on the color-breaking scale $w$.

The SM quarks are extended to fundamentals of SU(4$)_{cA}$. For instance up-type quarks are described as
\beq
\mathcal{U}=\left(\begin{array}{c}
  U \\
U_4
\end{array}\right)~,
\eeq
where $U$ is associated with SM up-type quarks and $U_4$ is a BSM color singlet. The down-type $D$ quarks and left-handed quarks $Q$ are extended in similar ways to $\mathcal{D}$ and $\mathcal{Q}$, respectively. The gauge quantum numbers of these multiplets are
\beq
\mathcal{D}\sim\left(\bm{4},\bm{1},-\frac{3}{8} \right)~, \ \ \mathcal{U}\sim\left(\bm{4},\bm{1},\frac{5}{8} \right)~, \ \ \mathcal{Q}\sim\left(\bm{4},\bm{2},\frac{1}{8} \right)~.\label{eq:Qcharges}
\eeq
These U(1$)_X$ charges are those for which all anomalies vanish and they give rise to the usual hypercharges of the SU(3$)_c$ quark fields. The Higgs, with gauge charges
\beq
H\sim\left(\bm{1},\bm{2},\frac{1}{2} \right)~,
\eeq
couples to these SU(4$)_c$ color multiplets in the expected extension of the SM Yukawa terms
\beq
\lambda_D\overline{\mathcal{Q}}H\mathcal{D}+\lambda_U\overline{\mathcal{Q}}\widetilde{H}\mathcal{U}+\text{H.c}~.
\eeq
Similar operators (with $A$ sector fields exchanged for their $B$ sector partners) appear in the twin sector.

If the mass of these new fermionic states only arose from electroweak symmetry breaking, then they would have been discovered long ago. Therefore, any viable model must include an explanation of how these fields obtain higher masses. We assume that in the visible sector there exist three generations of color singlet fields $\widehat{Q}$, $\widehat{U}$, and $\widehat{D}$ with quantum numbers
\beq
\label{eq:exotic-fermion}
\widehat{D}\sim\left(\bm{1},\bm{1},-\frac{1}{2} \right)~, \ \ \widehat{U}\sim\left(\bm{1},\bm{1},\frac{1}{2} \right)~, \ \ \widehat{Q}\sim\left(\bm{1},\bm{2},0 \right)~.
\eeq
Again, these U(1$)_X$ charges ensure that the model remains anomaly-free while including these fermions. These additional fermions couple to the SU(4$)_c$ fields through the Yukawa terms 
\beq
\lambda_{\widehat{Q}}\overline{\mathcal{Q}}\Phi \widehat{Q}+\lambda_{\widehat{U}}\overline{\mathcal{U}}\Phi \widehat{U}+\lambda_{\widehat{D}}\overline{\mathcal{D}}\Phi \widehat{D}+\text{H.c.}~.\label{e.QuarkBSMFerm}
\eeq
When $\Phi_A$ gets a VEV these interactions produce $w$-scale mass terms. Similarly to the SM quarks, the exact value of the masses can be much lower than the VEV if the corresponding Yukawa couplings are small. The equivalent interactions in the twin sector do not produce masses because $\Phi_B$ does not develop a VEV. The $\Phi_{A,B}$ fields do not carry baryon number, so these interactions preserve that symmetry by assigning $B=1/3$ to each of the hatted fields, the same value as the SM quarks. 

In addition to these mass terms, the gauge symmetries of the theory allow the additional Higgs interactions
\beq
\lambda_{\widehat{Q}\widehat{D}}\widehat{\overline{Q}}H\widehat{D}+\lambda_{\widehat{Q}\widehat{U}}\widehat{\overline{Q}}\widetilde{H}\widehat{U}~+\text{H.c.}~.\label{eq:hatMasses}
\eeq
These provide an electroweak-scale mass to the new fermions. Importantly, in the twin sector these masses ensure that the new fermionic states (which receive no color-breaking mass) are not massless. At low energies the mass eigenstates are Dirac fermions with baryon number 1/3. One state has twin electric charge of 1/2 while the other has twin electric charge of $-$1/2. 

As shown in Appendix~\ref{a.Details}, after SU(4$)_c$ color breaking the mass eigenstates of the BSM fermions have electric charge $\pm\frac12$. They can be pair-produced through neutral vector bosons or through the $t$-channel exchange of a $\xi^\mu$ from the interaction term
\beq
\frac{g_s}{\sqrt{2}}\left[\, \overline{U}\slashed{\xi}U_4+\overline{D}\slashed{\xi}D_4+\overline{Q}\slashed{\xi}Q_4 \right]+\text{ H.c.}~.\label{e.XiInt}
\eeq
There have been only a few LHC searches~\cite{CMS:2012xi,CMS:2024eyx,Vannerom:2019zsi} for fractionally charged particles. These searches robustly require these fermions to have masses above 600 GeV, and can likely be strengthened in the near future. Along with searches for the neutral $Z'_\mu$ and color charged vector $\xi_\mu$, these fractionally charged fermions make an interesting collider target for the LHC.

The interactions in Eq.~\eqref{e.XiInt} have the potential to produce flavor violating interactions observables like $K$-$\overline{K}$ mixing. However, if the fourth-component fields transform the same way under flavor as their colored counterparts then these interactions remain flavor diagonal. Because the fourth-component fields mix with the hatted fields through the mass term in~\eqref{e.QuarkBSMFerm} we also require the hatted field to carry the same flavor as their unhatted counterparts. In essence, this structure allows our model to realizes minimal flavor violation~\cite{DAmbrosio:2002vsn} (MFV), which is known to offer significant protection against new sources of flavor violation. The simplest implementation of MFV is to take the Yukawa matrices in Eq.~\eqref{e.QuarkBSMFerm} to be diagonal in the appropriate flavor space. Similarly, the couplings $\lambda_{\widehat{Q}\widehat{D}}$ and $\lambda_{\widehat{Q}\widehat{U}}$ in Eq.~\eqref{eq:hatMasses} are taken to be proportional to the SM Yukawas $\lambda_D$ and $\lambda_U$, respectively. However, these assumptions can also be modified, which could produce interesting flavor signals.

\section{Collider Phenomenology\label{sec:pheno}}
In this section we determine the existing constraints on the parameters of our model due to experimental collider searches for the new vectors and fermions that arise in the visible sector. We also discuss new discovery opportunities at the LHC for these new states. Many of the formulas related to these analyses are recorded in Appendix~\ref{a.Pheno}.

\subsection{$Z'$ Searches\label{ssec:ZpSearch}}
We begin with the new heavy vectors that appear after SU(4$)_c$ color breaks in the visible sector. The details regarding these states are given in Appendix~\ref{a.Vecs}. The heavy $Z'$ is neutral under all SM gauge groups, similar to the $Z$. Also like the $Z$, it can be produced at the LHC through Drell-Yan processes. Both ATLAS and CMS have searched for such new states through dilepton~\cite{ATLAS:2019erb,CMS:2021ctt} and dijet~\cite{CMS:2018mgb,CMS:2019gwf} resonances. 

\begin{figure}
    \centering
    \includegraphics[width=0.6\linewidth]{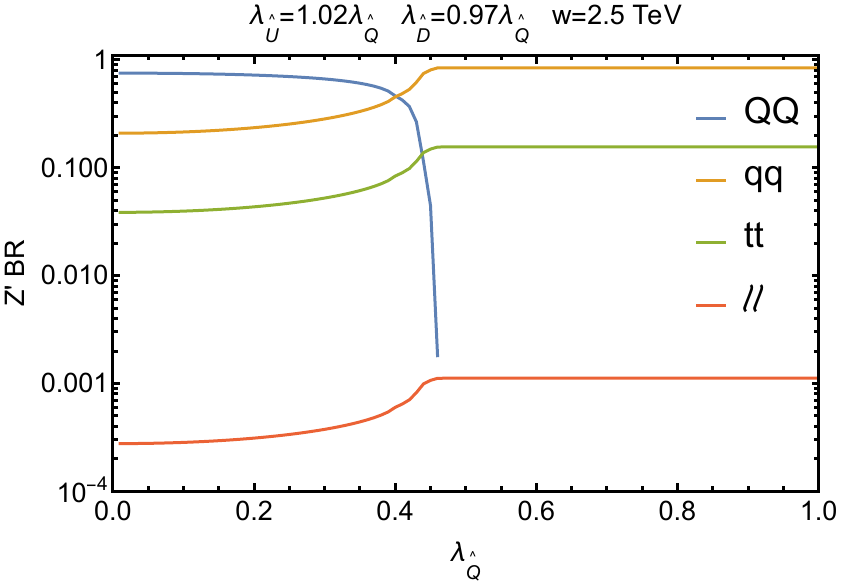}
    \caption{Dominant branching fractions of the heavy $Z'$ into BSM fermions $QQ$, SM quarks $qq$ (other than the top quark), $tt$, and SM leptons $\ell\ell$ (other than the tau). The color-breaking VEV is fixed at $w=$2.5 TeV while the Yukawa couplings of the BSM fermions to the color-breaking scalar vary, but are all taken to be of similar size. }
    \label{fig:ZpBRs}
\end{figure}

The $Z'$ arises from the SU(4$)_c$ color sector, so it has large couplings to the BSM fermions and SM quarks, with comparatively small couplings to leptons, as shown in Fig.~\ref{fig:ZpBRs}. In situations where the fourth-component quarks are kinematically available, the $Z'$ has branching fractions of about 21\% to SM quarks and 0.01\% to leptons. The use of quark, in this case, does not include the top quark. This is because di-jet searches at the LHC use the lighter five flavors in their analyses. Similarly, the term lepton should be understood as applying only to the electron and muon, as these are the states used in di-lepton resonance searches.

When the BSM fermions are too heavy for the $Z'$ to decay into them, the branching fraction into leptons goes up to about 0.1\% and the fraction into light SM quarks is about 84\%. The $Z'$ can also decay into $Zh$ and $WW$, but the branching fractions are too small for these channels to play a significant discovery role. The total decay width of the $Z'$ is also sensitive to whether it can decay to the fourth-component quarks. When they are kinetically available, the width of the $Z'$ is about 8\% of its mass. If the $Z'$ cannot decay into these BSM fermions, then its decay width is about 2\% of its mass. 

\begin{figure}
    \centering
    \includegraphics[width=0.49\textwidth]{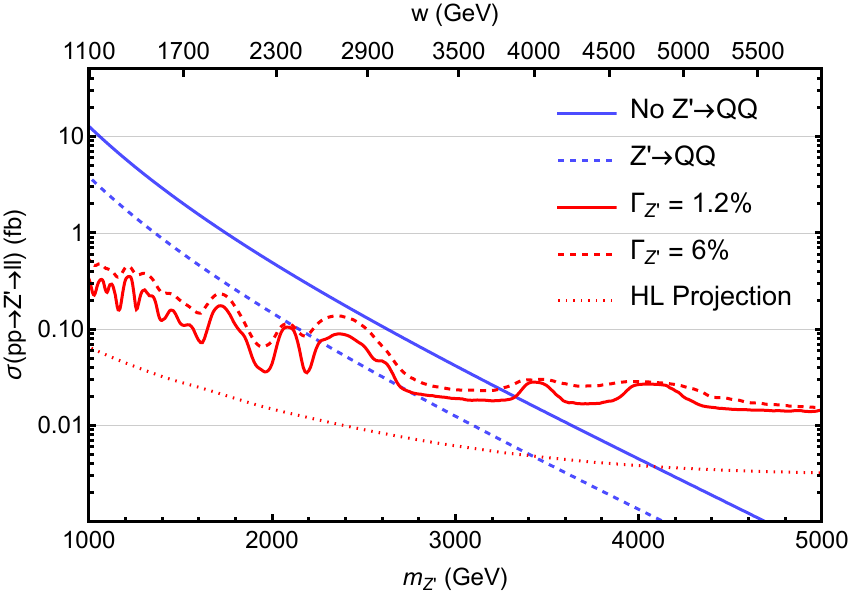}
    \includegraphics[width=0.49\textwidth]{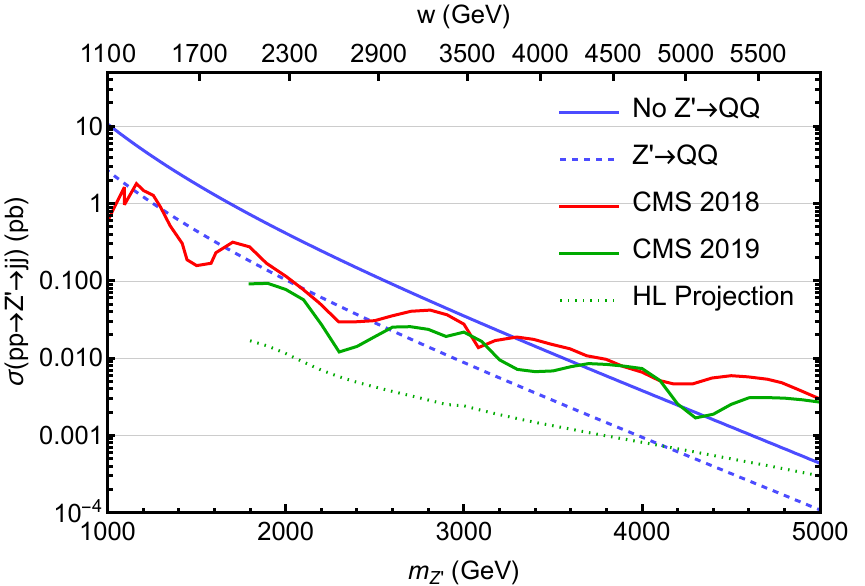}
    \caption{On the left (right) we plot the bounds, in red, on the $Z'$ from an ATLAS dilepton search~\cite{ATLAS:2019erb} (CMS dijet searches~\cite{CMS:2018mgb,CMS:2019gwf}). The solid (dashed) blue lines denote the cross section when $Z'$ decays to BSM fermions are (not) allowed. The bounds are given as a function of the $Z'$ mass and also as a function of the color-breaking VEV, $w$.}
    \label{f.ZpBounds}
\end{figure}

The sensitivity of the branching fractions to the fourth-component quark masses impact the corresponding sensitivities in LHC searches. In Fig.~\ref{f.ZpBounds} we see bounds on the $Z'$ mass from dilepton (left) and dijet (right) searches. In both plots the solid (dashed) blue line corresponds to the case when the $Z'$ cannot (can) decay into the BSM fermions, giving larger (smaller) branching into SM leptons and quarks. The experimental dilepton search results (on the left) are given in terms of the decay width of the $Z'$. The solid (dashed) red line denotes the bound for $\Gamma_{Z'}/m_{Z'}=0.012$ (0.06). The searches are more sensitive for smaller decay widths, so both of these limits provide a somewhat conservative estimate of the true bound. We see that if the $Z'$ can decay into the fourth-component quarks it must have a mass of a little more than 2 TeV, corresponding to a color-breaking VEV $w$ just below 2.5 TeV. When the BSM fermions are too heavy for the $Z'$ to decay into them the bound on the mass rises to nearly 3.5 TeV, or a $w$ of close to 4 TeV.

The dotted red line is a rough projection of the discovery sensitivity of the high luminosity LHC. This is obtained by rescaling the reported experimental cross sections by the square root of the ratio of the luminosities
\beq
\sigma_\text{HL}=\sigma_{13\text{TeV}}\sqrt{\frac{\mathcal{L}_{13\text{TeV}}}{\mathcal{L}_\text{HL}}}~,
\eeq
assuming 3000 $\text{fb}^{-1}$ of luminosity during the high-luminosity run. For the dilepton sensitivity we use the expected value of the $\Gamma_{Z'}/m_{Z'}=0.012$ bound.

The right panel of Fig.~\ref{f.ZpBounds} shows that the dijet searches are somewhat more powerful $Z'$ discovery tools in this scenario. Two CMS searches are shown in red and green. The latter bound is also projected to expected high-luminosity sensitivity. We find that bounds from these searches are similar to the dilepton sensitivity, but somewhat stronger. The high-luminosity projection indicates that even when the $Z'$ can decay into the BSM fermions, vector masses above 4 TeV (color-breaking scales of almost 5 TeV) will eventually be probed by the LHC. We also note that $t\overline{t}$ resonances searches~\cite{CMS:2018rkg,ATLAS:2020lks} can also be applied to this scenario, but the resulting bounds are somewhat weaker than the di-lepton searches.

\subsection{Colored Vectors with Fractional Electric Charge} 
The other SU(4$)_c$ gluon states that become massive are gathered into a colored vector field $\xi^\mu$. This vector can be pair produced through SM gluons or through any of the neutral vector bosons, $Z'$, $Z$, and $\gamma$. A $\xi^\mu$ can also be produced in association with a fourth-component quark field through its coupling to the SM quarks, see Eq.~\eqref{e.XiInt}. 

If these BSM fermions are lighter than the $\xi^\mu$ then these same interactions lead the the vectors decay promptly into the new fermion and a SM quark. As the mass of the $\xi^\mu$, see Eq.~\eqref{eq:mXiMBprime}, is approximately the same as the $Z'$ we are primarily interested in cases with the new fermions significantly lighter than the $\xi^\mu$. For completeness, however, we do make some discussion of these vectors when they cannot decay.

The $\xi^\mu$ has electric charge of $1/6$, which makes its phenomenology quite interesting. It cannot decay into any SM states. Because it carries color, after production this state would shower and hadronize, eventually becoming bound to a light SM quark. This would be an exotic bound state with electric charge of $\pm1/2$.

Collider searches for states with fractional charge are, surprisingly, scarce. However, the recently growing literature~\cite{Alonso:2024pmq,Li:2024nuo,Koren:2024xof,Koren:2025utp} motivating and exploring such states will hopefully stimulate further exploration. We note the distinction from millicharged particle searches, which consider very small electric charges, compared with the order one, but less than one, electric charges. Results from LEP data~\cite{OPAL:2003zpa} have been used~\cite{Davidson:2000hf} to exclude particles with charge greater than 2/3 and masses below 95 GeV. Early direct searches from CMS~\cite{CMS:2012xi,CMS:2013czn} have also focused on charge 2/3 states, reaching bounds of about 500 GeV. A recent CMS search~\cite{CMS:2024eyx} considers a wider range of possible electric charges. For the charge 1/2 possibilities that are most relevant to this work, the published bound is 600 GeV. 

This bound assumes that production only proceeds through the electric coupling, so it does not immediately apply to the $\xi^\mu$ bound states. To date, however, more general bounds in terms of a possibly larger cross section have not been produced.\footnote{Plots of this type do appear in~\cite{Vannerom:2019zsi}, but there seems to have been an unrecognized issue regarding tracking in the muon system which affects the sensitivity for lower charges. We are very grateful to Steven Lowette for clarifying the situation.} So, while it may well be that the LHC sensitivity to other types of fractionally charged particles is significantly greater than 600 GeV, we cannot yet apply a more definite bound.

The upshot is that novel bound states with electric charge of 1/2 and masses near 2 TeV are completely unconstrained by current searches. Therefore, the $\xi^\mu$ vector is unlikely to provide discovery opportunities until well after other states, like the $Z'$, have been detected. In fact, fractionally charged bound states containing $\xi^\mu$ would likely be discovered after the other, elementary, fractionally charged particles that appear in our framework.

\subsection{Fractionally Charged Fermions}
\begin{figure}
    \centering
    \includegraphics[width=0.8\linewidth]{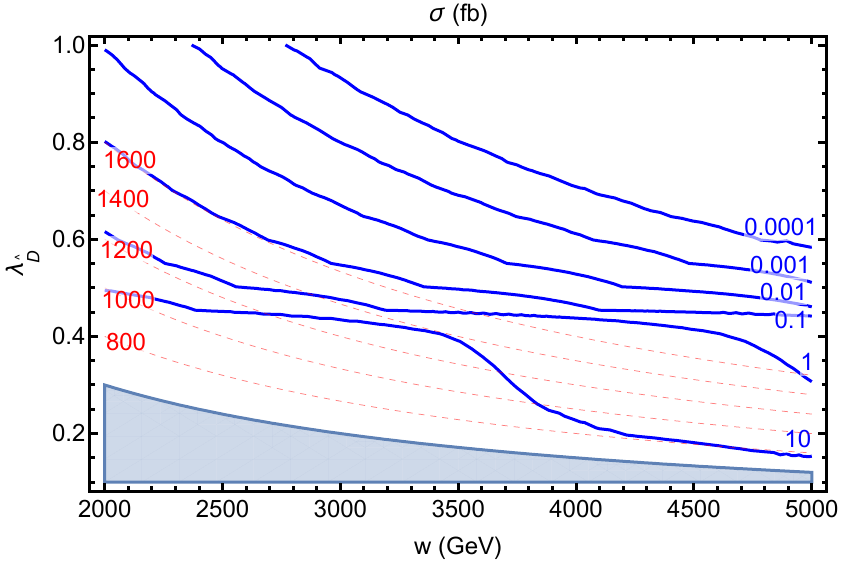}
    \caption{13 TeV LHC production cross section for down-quark type fourth-component quarks given in blue. The shaded region is excluded by fractionally charged particle searches. Dashed red lines denote values of $\lambda_{\widehat{D}} w$, approximately the fermion mass.}
    \label{fig:DDQproduction}
\end{figure}

As detailed in Appendix~\ref{sa.bsmFerm}, after SU($4)_c$ color breaking, the fourth components of the quark fields appear as novel charged 1/2 fermionic states. They can be produced through $s$-channel neutral bosons $\gamma,\,Z,\,$ and $Z'$ or through the exchange of a $t$-channel $\xi$. See Appendix~\ref{a.Pheno} for the relevant cross sections. The combined cross section from these two channels for the down-type fermions (to all three flavors) is shown as blue contours in Fig.~\ref{fig:DDQproduction}. The production of the up-type states is qualitatively similar. 

In the figure, red dashed curves of approximate fermion mass, $\lambda_{\widehat{D}} w$, are overlayed for reference. One can see that the production is enhanced when the intermediate, $s$-channel $Z'$ can go on-shell. That is, when raising $w$ not only increases the mass of the fermions, but also makes it so that $m_{Z'}$ is less than twice fermion mass, the cross section is larger. We also shade the published bound on fractionally charged particles. While we have only shown the bound for the down-type states, they are qualitatively similar to each of the fourth-component fermions.

The cross section for these BSM fermions is greater than what would result from an $s$-channel photon and $Z$ alone, so the bound from the CMS study~\cite{CMS:2024eyx} is likely greater than 600 GeV. As with the $\xi^\mu$, however, we cannot reliably estimate how the bound changes for a larger cross section. Future LHC analyses are likely to probe these states to much higher masses. In other words, this type of extended color model provides a definite target for fractionally charged particle searches.

For definiteness in this study we simply consider the benchmarks of 600 GeV and 1 TeV to illustrate possible phenomenologies. We note that for both of these masses the fourth-component quarks are less than half the $Z'$ mass when it comes to searches for this new vector. Therefore, the di-jet and di-lepton resonance searches are less sensitive and the bound on $w$ is lower.

\begin{figure}
    \centering
    \includegraphics[width=0.49\linewidth]{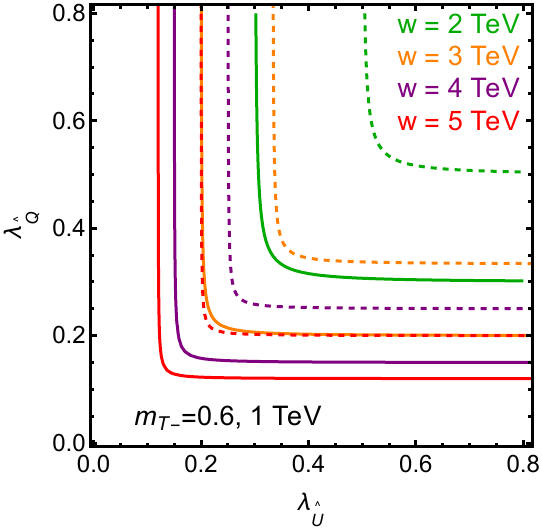}
    \includegraphics[width=0.49\linewidth]{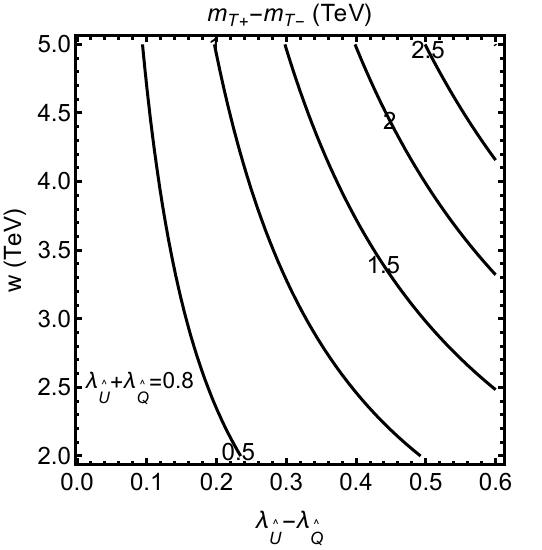}
    \caption{Left: Yukawa couplings for the lower mass, fourth-component, top-type, fermions with masses of 600 GeV (solid) and 1 TeV (dashed). These are given for several values of $w$. Right: Mass splitting between the two top-type states as a function of $w$ and $\lambda_{\widehat{U}}-\lambda_{\widehat{Q}}$.}
    \label{fig:fermYukawas}
\end{figure}

In Fig.~\ref{fig:fermYukawas} we show the possible values of the Yukawa couplings of the SU(4$)_c$ quark fields to the color-breaking scalar $\Phi_A$. The left panel corresponds to the lighter top-type state, $T_{-}$. The 600 GeV mass benchmark is denoted with solid lines while the 1 TeV benchmark uses dashed lines. Several different values of the color-breaking VEV $w$ are shown. 

We see that if we require the $T_{-}$ state to have at least a certain mass, then both of the Yukawa couplings must be sufficiently large. For example, if $w=3$ TeV and the fermion is to have a 600 GeV mass, then $\lambda_{\widehat{U}}$ and $\lambda_{\widehat{Q}}$ must have a value of at least 0.2 . In contrast, imposing a similar mass requirement on the lighter state requires both of the Yukawa couplings to be larger than some value.  

In the right panel of Fig.~\ref{fig:fermYukawas} we show how the mass splitting between the two top-like states depends on the Yukawa couplings and color breaking scale. The spitting is insensitive to the sum of the Yukawas, but grows with their difference. We have assumed in the plot that $\lambda_{\widehat{U}}>\lambda_{\widehat{Q}}$, but the opposite assumption leads to identical results.

\section{Scalar Potential\label{sec:Higgs}}
In this section we describe the scalar potential of the extended color twin Higgs model. We also examine the tuning related to varying the potential and other model parameters and compare this to the tuning in the standard twin Higgs construction. For simplicity, we restrict our analysis to the tree-level potential in our discussion here; however, one loop contributions can be important.  These contributions to the potential, along with many of the computational details regarding the tuning, can be found in Appendix~\ref{ap:tunig}.

The scalars that appear in the potential include the SU(2$)_L$ doublet Higgs fields $H_{A,B}$ of each sector and the scalar fundamentals of SU(4$)_c$ color $\Phi_{A,B}$. The tree level potential for all these scalars is given by
\begin{align}
    V =& -M_H^2 |H|^2+\lambda_H |H|^4 -M_\Phi^2 |\Phi|^2+\lambda_\Phi |\Phi|^4+\lambda_{H \Phi}|H|^2 |\Phi|^2\\ 
    &+\delta_H (|H_A|^4+|H_B|^4)
    +\delta_\Phi\left(|\Phi_A|^4+|\Phi_B|^4\right)+ \delta_{H\Phi}(|H_A|^2-|H_B|^2)\left(|\Phi_A|^2-|\Phi_B|^2\right)\nonumber~,
\end{align}
where
\beq
H=\left( \begin{array}{c}
    H_A \\
    H_B
\end{array}\right)~, \ \ \text{and} \ \ \Phi=\left( \begin{array}{c}
    \Phi_A \\
    \Phi_B
\end{array}\right)~.
\eeq
The terms in the top line of $V$ exhibit a large SU(4)$\times$SU(8) global symmetry. The SU(4) follows from the two Higgs fields and the SU(8) from the two $\Phi$ fields. The second line of the potential breaks these global symmetries but does preserve the discrete twin symmetry that exchanges $A$ and $B$ sector fields.

Because the low energy physics of the visible sector must exhibit SU(3$)_c$ symmetry we are interested in the scenario in which $\langle\Phi_A\rangle=w>0$ and $\langle \Phi_B\rangle=0$. It is also convenient to parameterize the Higgs VEVs by $\langle H_A\rangle=v_A/\sqrt{2}$ and $\langle H_B\rangle=v_B/\sqrt{2}$. It is shown in~\cite{Barbieri:2005ri} that without the $\Phi$ field and for $\delta_H>0$ the vacuum preserves the $Z_2$ symmetry, so $v_A=v_B$. Similarly, in the absence of $H$ and for $\delta_\Phi<0$ the vacuum spontaneously breaks the $Z_2$ symmetry such that all of the $\Phi$ VEV is in one of $\Phi_A$ or $\Phi_B$. When all the fields are included there remain parameter ranges that produce the vacuum structure we are interested in, where the color breaking occurs only in the visible sector and Higgs VEV is shared between both sectors.

Similar to the constructions in~\cite{Batell:2019ptb,Batell:2020qad}, the full potential with $\delta_H>0$ and $\delta_\Phi<0$ provides the vacuum structure we wish to explore. All of the $\Phi$ VEV can be in $\Phi_A$ while $\langle\Phi_B\rangle=0$. This produces an effective $Z_2$-
breaking mass term for $H_A$ and $H_B$ which drives $v_B>v_A$, as required for viable twin Higgs phenomenology. The degree of splitting between the two Higgs VEVs is controlled by the size of this mass term, $\delta_{H\Phi}w^2$.

To understand how the potential parameters relate to physical observables we use a nonlinear parameterization of the Higgs fields (similar to that of~\cite{Burdman:2014zta}) such that
\begin{align}
H_A=&\,\bm{h}\frac{f}{\sqrt{\bm{h}^\dag\bm{h}}}\sin\left(\frac{\sqrt{\bm{h}^\dag\bm{h}}}{f} \right)~,&
H_B=&\,\left(\begin{array}{c}
    0 \\
    \displaystyle f\cos\left(\frac{\sqrt{\bm{h}^\dag\bm{h}}}{f} \right) 
\end{array} \right)~,
\end{align}
where $f$ is the scale at which the global SU(4) symmetry of the Higgs and twin Higgs potential is spontaneously broken. Similarly, the $\Phi$ fields are taken to be 
\beq
\label{eq:Phi-nonlinear}
\Phi_A=\left(\begin{array}{c}
0\\
0\\
0\\
\displaystyle w\cos\left(\frac{\sqrt{|\phi_B|^2}}{w}\right)
\end{array}
\right)~, \ \ \ \ \Phi_B=\phi_B\frac{w}{\sqrt{|\phi_B|^2}}\sin\left(\frac{\sqrt{|\phi_B|^2}}{w}\right)~.
\eeq

Using this parameterization of the fields, and dropping constant terms, the potential is 
\begin{align}
V=&-\frac{\delta_Hf^4}{2} \sin^2\left(\frac{2\sqrt{\bm{h}^\dag\bm{h}}}{f} \right)-\frac{\delta_\Phi w^4}{2}\sin^2\left(\frac{2\sqrt{|\phi_B|^2}}{w}\right)\nonumber\\
&-\delta_{H\Phi}w^2f^2\cos\left(\frac{2\sqrt{\bm{h}^\dag\bm{h}}}{f} \right)\cos\left(\frac{2\sqrt{|\phi_B|^2}}{w}\right)~.
\end{align}
The Higgs potential is obtained by taking $\phi_B$ to its zero VEV 
\begin{align}
    V=&-\frac{\delta_Hf^4}{2}\sin^2\left(\frac{2\sqrt{\bm{h}^\dag\bm{h}}}{f} \right)-\delta_{H\Phi}w^2f^2\cos\left(\frac{2\sqrt{\bm{h}^\dag\bm{h}}}{f} \right)\nonumber\\
    &=2\bm{h}^\dag\bm{h}\left(w^2\delta_{H\Phi}-f^2\delta_H \right)+\frac23\left( \bm{h}^\dag\bm{h}\right)^2\left( 4\delta_H-\frac{w^2}{f^2}\delta_{H\Phi}\right)+\ldots~.
\end{align}
The second line illustrates how the potential parameters can be related to the quantities in the SM Higgs potential
\beq
V_\text{SM}=-\mu^2\bm{h}^\dag\bm{h}+\lambda\left( \bm{h}^\dag\bm{h}\right)^2~.
\eeq

When $\bm{h}$ is set to its VEV
\beq
\bm{h}\to
\left(\begin{array}{c}
    0 \\
  \displaystyle  \frac{v}{\sqrt{2}}
    \end{array}\right)~,
\eeq
we find
\beq
v_A=f\sqrt{2}\sin\frac{v}{f\sqrt{2}}~, \ \ \ \ v_B=f\sqrt{2}\cos\frac{v}{f\sqrt{2}}~,
\eeq
which motivates defining the angle $\vartheta=v/(f\sqrt{2})$. The condition that this VEV is the minimum of the potential is
\beq
w^2\delta_{H\Phi}=f^2\delta_H\cos(2\vartheta)~.\label{eq:vacVEV}
\eeq

We note that the fermion masses that arise only from electroweak symmetry breaking satisfy
\beq
\lambda_i^2f^2=\frac{\lambda_i^2}{2}\left(v_A^2+v_B^2 \right)=m_{iA}^2+m_{iB}^2~.
\eeq
That is, the scale $f$ is shared between the $A$ and $B$ sectors. The smaller $\vartheta$ the larger the twin particle masses as compared with the visible sector. As shown in~\cite{Burdman:2014zta}, the coupling of $\bm{h}$ to SM fields carries a factor of $\cos\vartheta$ compared to the SM prediction. Consequently, Higgs coupling measurements require that $\vartheta$ must be small. This predicts larger twin sector masses, which is associated with larger tuning. Therefore, the most natural realizations of the twin Higgs idea keep $\vartheta$ as large as experiment allows, implying that $f$ is only a factor of a few larger than $v_A$. 

In contrast, in Sec.~\ref{sec:pheno} we found that the color-breaking VEV must be greater than 2.5~TeV, so $w>f$. The relationship in Eq.~\eqref{eq:vacVEV} then implies that $0<\delta_{H\Phi}<\delta_H$. This gives us the schematic form of the scalar potential that gives rise to the scenario we are interested in. In fact, this argument can be further sharpened, as shown below.

By expanding about the vacuum $v\to v+h$, the mass of the Higgs boson is found to be
\beq
m_h^2=2f^2\delta_H\sin^22\vartheta~.
\eeq
This result can then be solved for $\delta_H$, yielding
\beq
\delta_H=\frac{m_h^2}{4v_A^2}\left(1+\frac{m_{tA}^2}{m_{tB}^2} \right)~.
\eeq
Using the known values of the Higgs and top quark masses as well as the electroweak VEV $v_A\approx246$ GeV we find 
\beq
\label{eq:deltaH-numerical}
\delta_H\simeq 0.064+7.7\times 10^{-3}\left(\frac{500\text{ GeV}}{m_{tB}}\right)^2~.
\eeq
Using this result, the vacuum relation in Eq.~\eqref{eq:vacVEV} leads to 
\beq
\label{eq:deltaHPhi-numerical}
\delta_{H\Phi}=\frac{m_h^2m_{tB}^2}{8w^2m_{tA}^2}\left(1-\frac{m_{tA}^4}{m_{tB}^4} \right)\simeq1.8\times10^{-3}\left( \frac{3\text{ TeV}}{w}\right)^2 \left(\frac{\phantom{I}m_{tB}\phantom{I}}{500\text{ GeV}}\right)^2~.
\eeq
Note that when the two top masses are equal that this coupling vanishes. This is simply the manifestation of this coupling (and its sign) determining the hierarchy of VEVs in the visible and twin sectors. We see that both $\delta_H$ and $\delta_{H\Phi}$, which break the global SU(4) symmetry of the Higgs potential, are small. This is what is required for the the Higgs boson $h$ to be a pNGB of the approximate global symmetry and have a mass well below the scale $f$.

To streamline the discussion, our analysis above has considered only the tree-level potential. However, one-loop contributions to the potential arising from the exchange of top quark sector states are numerically important. For example, these can yield corrections to the effective quartic couplings of similar quantitative size to the estimates in Eqs.~(\ref{eq:deltaH-numerical}) and (\ref{eq:deltaHPhi-numerical}). Furthermore, incorporating radiative corrections to the potential are essential for a proper assessment of the tuning, which we now discuss. Appendix~\ref{ap:tunig} contains  further technical details related to these issues.

\subsection{Tuning\label{ssec:tuning}}
One virtue of the twin Higgs framework is that it addresses the little hierarchy problem without introducing symmetry partners to the top quark with SM color charge. In the extended color variation of the twin Higgs one might worry that the tuning is increased because 
the fourth color component of the top quark's SU(4$)_c$ multiplet has a large mass, in addition to the twin sector states. Interestingly, our analysis reveals that the tuning is often comparable to, and in some cases even milder than, the standard mirror twin Higgs. This stems from structural features of the model that enforce partial cancellations among the various sources contributing to the Higgs squared mass parameter.

To make our tuning analysis more concrete we employ the measure suggested by Barbieri and Giudice~\cite{Barbieri:1987fn}. They define the tuning of a quantity $M$ due to the variation of a parameter $\lambda$ to be
\beq
\Delta_\lambda=\left|\frac{\lambda}{M}\frac{\partial M}{\partial \lambda} \right|~.
\eeq
Often, $M$ depends on several parameters. The reported tuning associated with the parameter is simply taken to be the maximum of the individual tunings. It is also usual to speak of the percent tuning of a parameters. This is simply obtained by inverting the Barbieri-Giudice value. That is, a $\Delta_\lambda$ of 10 corresponds to $\lambda$ being tuned to 10\%.  

In Appendix~\ref{ap:tunig} we calculate the tuning in the Higgs mass parameter $\mu^2$ for both the mirror twin Higgs and extended color twin Higgs constructions so that we can compare the two. Our results include both the tree-level and one-loop contributions to the scalar potential by using the results of Coleman and Weinberg~\cite{Coleman:1973jx}. The resulting equations are somewhat lengthy, so their full form is relegated to the appendix, but the qualitative implications are relatively straightforward to understand.

We first discuss the standard mirror twin Higgs. For nearly all parameters the tuning takes the form
\beq
\label{eq:MTH-tuning}
\Delta_\text{MTH}\sim\frac{f^2}{v^2}-\frac{3\lambda_t^4f^2}{4\pi^2m_h^2}\ln\frac{f}{v}~,
\eeq
as shown in Eqs.~\eqref{eq:THmTune}\textendash\eqref{eq:THf2TuneExp}.
This result features the familiar $f^2/v^2$ tuning arising at tree-level along with a subleading loop level contribution which, importantly, partially cancels the leading term. This partial cancellation is unavoidable and follows from the structure of the mirror twin Higgs model. 

Next, we consider the tuning in the extended color twin Higgs model. There are two important qualitative effects which help to mitigate the tuning in this scenario. First, the fact that the number of colors is 4 rather instead of 3 leads to a stronger partial cancellation between the tree-level $f^2/v^2$ tuning and the loop-level one associated with certain parameters. For example, the tuning associated with the parameter $f^2$, which dominates over much of the parameter space, is given by
\beq
\label{eq:fsq-tuning}
\Delta_{f^2}\sim\frac{f^2}{v^2}-\frac{4\lambda_t^4f^2}{4\pi^2m_h^2}\ln\frac{f}{v}~.
\eeq
This equation can be compared to the Eq.~(\ref{eq:MTH-tuning}) for the standard mirror twin Higgs.

\begin{figure}
    \centering
\includegraphics[width=0.8\textwidth]{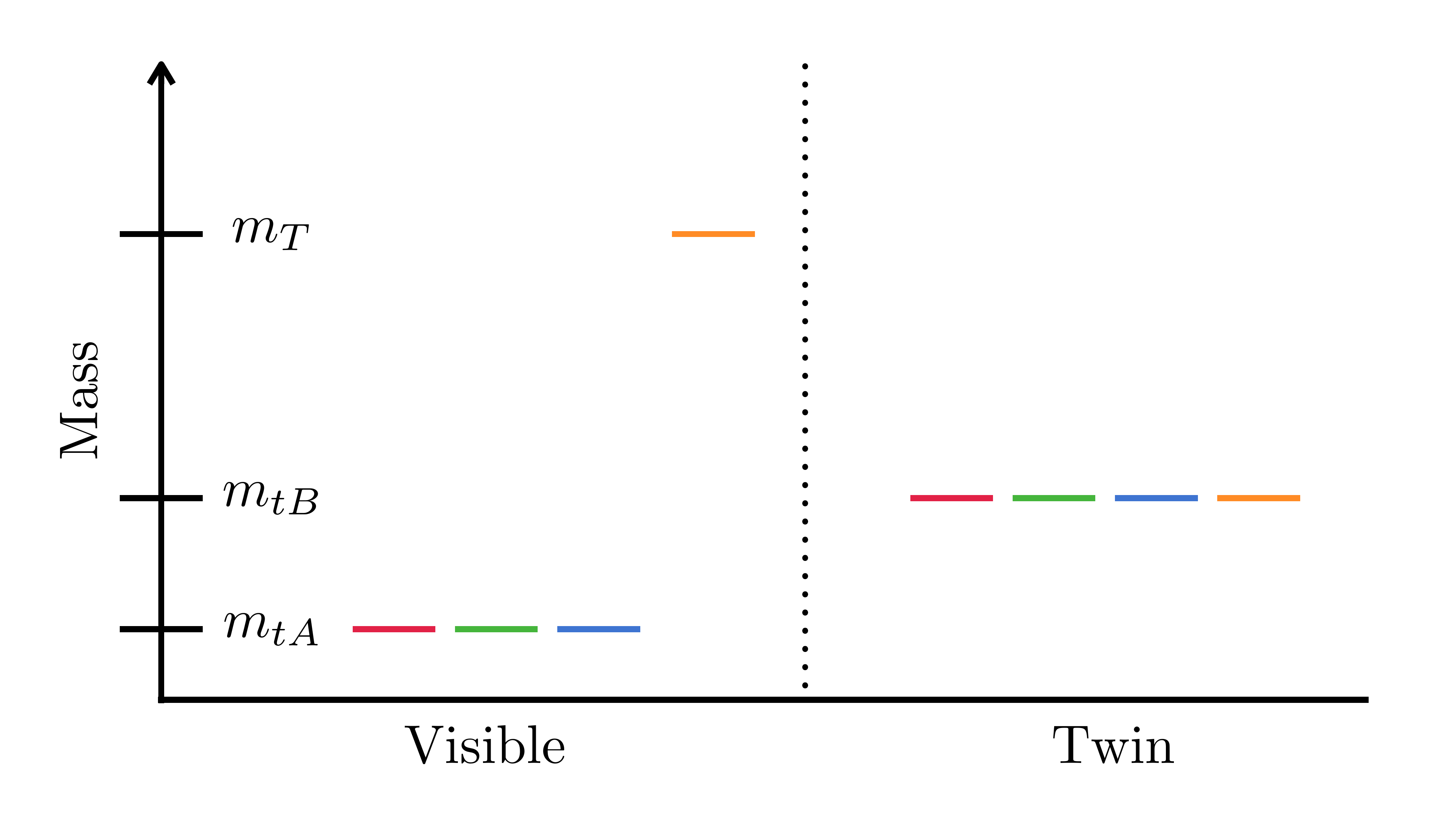}
    \caption{Schematic spectrum of the top-Yukawa states. SU(4$)_c$ color is broken in the visible sector and the fourth component of the top quark multiplet is raised to a high scale. In the twin sector color is unbroken, so all four fields have the same mass.}
    \label{f.spectrum}
\end{figure}

The second consequence comes from an even stronger cancellation in the tuning associated with certain parameters due to the spectrum of top quark states, see Fig.~\ref{f.spectrum}. One can see, calculationally, how the approximate SU(4) symmetry of the Higgs potential prevents large corrections to the Higgs mass parameter at one loop. The effect of the top quark loop in the visible sector is largely canceled by the contribution of the twin-top. In the typical twin Higgs set-up the portions of these loops that are quadratically sensitive to high scales enter with equal magnitude
\beq
\sim\frac{3\lambda_t^2\Lambda^2}{8\pi^2}~,\label{e.topLoop}
\eeq
and opposite sign. Thus, the three top quark colors cancel against the three twin-top colors. The remaining result depends on the log of $m_{tA}/m_{tB}$. When this ratio is large it makes a significant contribution to the tuning.

In the extended color twin Higgs the SM top-loop produces the usual term shown in~\eqref{e.topLoop}. The twin-top loop, however, produces
\beq
\sim-\frac{4\lambda_t^2\Lambda^2}{8\pi^2}~,
\eeq
because there are four twin-top states. Therefore, in contrast the standard twin Higgs, the twin states overcompensate for the top-loop. When the higher mass fourth-component top field is included the quadratic sensitivity to $\Lambda$ is removed.

 A similar interplay between the visible and twin sectors also applies to the subleading loop-level log terms. So, while both the additional visible top state and the twin states would both naively lead to more tuning, the structure of the theory is such that each partially cancels the effect of the other. For 
 certain parameter regions this cancellation can be nearly complete. As shown below, this happens to occur in the parameter space typically considered in twin Higgs models.

\begin{figure}
    \centering
    \includegraphics[width=0.85\textwidth]{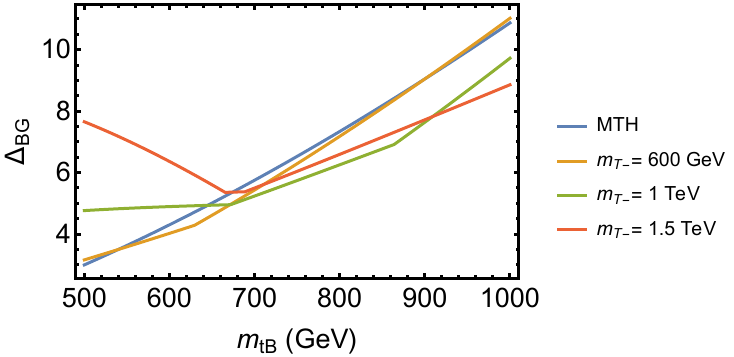}
    \caption{ Tuning in standard the mirror twin Higgs (blue) framework compared to tuning in the the extended color twin Higgs as a function of the twin top quark mass taking the cutoff $\Lambda=5$ TeV and the color breaking VEV $w=$ 2.5 TeV. The comparison is made for three values of the mass $m_{T_-}$ of the lightest fourth-component top quark. The masses of the heavier quark state have a much smaller affect upon the tuning. In this plot the values $m_{T_+}$ are 180 GeV larger than $m_{T_-}$.}
    \label{f.mTPlotTuning}
\end{figure}

Figure~\ref{f.mTPlotTuning} illustrates the results of these calculations. We plot the tuning in the mirror twin Higgs model in blue as a function of the twin top mass for a UV cutoff of $\Lambda=$ 5 TeV. This shows the familiar result that as the twin top mass increases (or as $\vartheta$ decreases) the tuning of the model increases. The curve is smooth because nearly all the parameters lead to the same form of the tuning given in Eq.~(\ref{eq:MTH-tuning}) above.

This outcome is compared with the tuning in the extended color model. The curves shown in tan, green, and red correspond to the lightest of the fourth-component top quark states having a mass $m_{T_-}$ of 600 GeV, 1 TeV, and 1.5 TeV, respectively. For the plot we choose $w=2.5$ TeV, but it is the actual physical masses that affect the tuning so the plot does illustrate the general result. 
For each case we have taken $m_{T_+}$ to be 180 GeV heavier than $m_{T_-}$, which keeps both of the states at essentially the same scale.
Unlike the mirror twin Higgs case, each of these curves has a kink where the maximum tuning changes from one parameter to another. For $m_{T_-}=600$ GeV, at lower $m_{tB}$ the maximum tuning is from $\Delta_{f^2}$, but this switches to $\Delta_{w^2}$ at about $m_{tB}\approx 600$ GeV. 
In contrast, when $m_{T_-} = 1$ TeV the tuning is initially dominated by $\Delta_{\lambda_{\widehat{U}}}$ at low $m_{tB}$, with $\Delta_{f^2}$ taking over near $m_{tB}\approx 700$ GeV, and eventually $\Delta_{w^2}$ overtakes the others around $m_{tB}\approx 900$ GeV.
Finally, when $m_{T_-}=1.5$ TeV the tuning at low $m_{tB}$ is larger and due to $\Delta_{w^2}$. However, this tuning decreases steeply as $m_{tB}$ increases due to the strong cancellation between different top sector states (see discussion above) until $\Delta_{f^2}$ becomes the largest tuning around $m_{tB}\approx 700$ GeV. 
Notice also that in each of the regions where $\Delta_{f^2}$ dominates, the tuning associated with this parameter is less than the tuning in the standard mirror twin Higgs due to the stronger cancellation between tree and loop contributions, a consequence of the number of colors being 4 instead of 3. 

These results show that the tuning is more intricate in the extended color model. When $m_{tB}$ is small but $m_{T_-}$ is large there is significant tuning. However, as $m_{tB}$ increases this tuning is reduced as the effects from the two sectors cancel one another. The significant conclusion is that for much of the motivated $m_{tB}$ parameter space the tuning in the extended color model is a little lower than the original twin Higgs.

\section{Cosmology\label{sec:cosmo}}
In this section we discuss some of the cosmological implications of this extended color twin Higgs framework. We first note that the bounds on charged $1/2$ relics are substantial~\cite{Perl:2001xi,Taoso:2007qk,Dunsky:2018mqs}. Therefore, any viable realization of this model must include a reheating temperature that is lower than the fourth-component fermion masses. Were these states to be produced in the early Universe then, because they cannot decay to SM states, that population would remain to the present. A reheating temperature that is lower than ${\cal O}$(100 GeV) ensures that these states are not produced and still allows for successful nucleosynthesis.\footnote{The bound on the reheating temperature may be more stringent if the maximum temperature following inflation is significantly larger than the reheating temperature~\cite{Giudice:2000ex}. See Ref.~\cite{Batell:2024idg} for a recent discussion in the context of fractionally charged particles.} A low reheating temperature would also serve to evade a potential domain wall problem associated with the spontaneous breaking of $Z_2$ symmetry. As is well known, low reheating temperatures pose significant challenges for baryogenesis.  However, it is possible to realize viable low-scale baryogenesis in the Twin Higgs framework through a number of mechansims, as shown  in Refs.~\cite{Farina:2016ndq,Earl:2019wjw,Feng:2020urb,Kilic:2021zqu,Beauchesne:2021opx,Bittar:2023kdl,Alonso-Alvarez:2023bat}. 

Without additional structure, the minimal mirror twin Higgs set-up is known to produce significant contributions to $\Delta N_\text{eff}$. The coupling of the Higgs to both visible $(A)$ and twin $(B)$ fermions keeps the sectors in thermal equilibrium until the Universe reaches the decoupling temperature $T_D\approx$~2 GeV for $f/v\sim4$ and increases to about 6 GeV for $f/v\sim10$~\cite{Craig:2016lyx}. As argued in~\cite{Chacko:2016hvu}, one can estimate the contribution to $\Delta N_\text{eff}$ by
\beq
\Delta N_\text{eff}\approx 7.4\left(\left.\frac{g_{\ast B}}{g_{\ast A}} \right|_{T_D} \right)^{4/3}~,
\eeq
where $g_\ast$ is the number of relativistic degrees of freedom
\beq
g_\ast=\sum\left(N_B+\frac78N_f \right)~.
\eeq
Here $N_B$ and $N_F$ are the number of boson and fermion (respectively) degrees of freedom with masses less than the temperature.

In the visible sector, $g_{\ast A}$ at $T_D$ is given by
\beq
g_{\ast A}=2\times 8+2+\frac78\left(3\times 4\times 4+3\times 4+3\times 2 \right)=75.75~,
\eeq
where we have included the gluon and photon fields, as well as four quark flavors, and all leptons. In the mirror twin Higgs scenario the fermion masses if increased by a factor of $\cot\vartheta$
\begin{equation}
m_{q B} = m_{q  A} \cot \vartheta \gtrsim 3  m_{q  A}~.
\end{equation}
Measurements of Higgs couplings imply that this value is greater than three. Tuning considerations favor a value of $\cot\vartheta$ not much greater than about six. For values in this range
\beq
g_{\ast B}=2\times 8+2+\frac78\left(3\times 4\times 3+2\times 4+3\times 2 \right)=61.75~,
\eeq
because the twin charm and tau do not contribute. This leads to
\beq
\Delta N_\text{eff}\approx 5.7~,\label{eq:NeffTH}
\eeq
which is well beyond the cosmological bounds~\cite{Planck:2018vyg}.

There are many ways that this simple story can be modified and produce contributions to $\Delta N_\text{eff}$ that would not yet have been detected. These include various ways to induce asymmetric reheating~\cite{Farina:2015uea,Chacko:2016hvu,Csaki:2017spo,Liu:2019ixm,Beauchesne:2021opx,Bittar:2024ryj} and modifying the decoupling temperature~\cite{Harigaya:2019shz}. These modifications of the twin Higgs framework are often quite modular and could easily be included in the color-breaking scenario that is the focus of this work.

\begin{figure}
    \centering
    \includegraphics[width=0.8\linewidth]{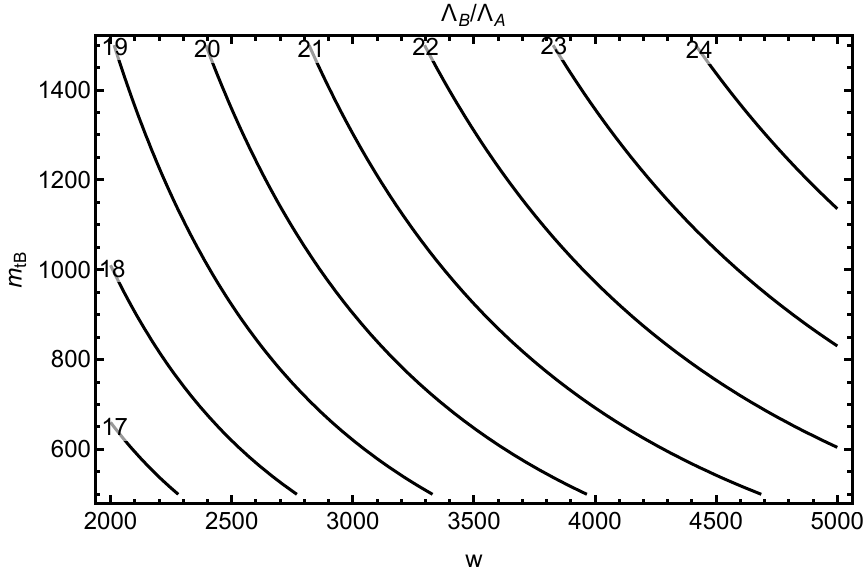}
    \caption{Contours of the two-loop estimate of the ratio of twin sector confinement scale $\Lambda_B$ to the visible sector confinement scale $\Lambda_A$ as a function of the twin top mass and the color-breaking VEV.}
    \label{fig:confinementScale}
\end{figure}

In the present model the simple cosmological history presented above is modified in one very significant way. In the twin sector, where the SU($4)_c$ color remains unbroken, the larger number of massless gluons leads to a higher confinement scale, see Fig.~\ref{fig:confinementScale}. This scale is estimated by taking the strong couplings in both sectors to be equal (due to the $Z_2$ symmetry) at the color-breaking scale $w$. We then use two-loop running  of the twin coupling to estimate the confinement scale, see appendix~\ref{ap:Running} for more details. The most significant aspect of these possible confinement scales is that they are all larger than the temperature at which the two sectors decouple. 

One consequence of this higher confinement scale is that the twin sector undergoes its QCD phase transition while the two sectors are still in equilibrium. The heat produced during the phase transition is consequently shared among the two sectors. As the Universe cools the two sectors equilibrate and then decouple. The visible sector then undergoes its QCD phase transition, which only heats the SM fields. This difference in temperature suppresses the contributions of light twin sector fields to $\Delta N_\text{eff}$. 

After the twin QCD phase transition, $g_{\ast B}$ is significantly reduced. The meson spectrum of the twin sector may be somewhat different than those measured in the visible sector because now even the twin charm has a mass below the confinement scale, though not much below. In any case the usual analysis of chiral symmetry breaking indicates the presence of light pions (see below for further details). If these twin pions are still present in the thermal bath at the time of decoupling then
\beq
g_{\ast B}=2+3\times 1+\frac78\left(2\times 4+3\times2 \right)=17.25~.
\eeq
If the twin pion masses are above the decoupling temperature then 
\beq
g_{\ast B}=2+\frac78\left(2\times 4+3\times2 \right)=14.25~.
\eeq
The implication is that for our extended color version of the twin Higgs framework that without considering any other mechanisms 
\beq
N_\text{eff}\approx0.8\text{\textendash}1.0~.
\eeq
While this still exceeds the bounds from the CMB~\cite{Planck:2018vyg}, we see that the contribution to $\Delta N_\text{eff}$ has been significantly reduced compared to the usual twin Higgs result~\eqref{eq:NeffTH}. We also note that this process is similar to the method suggested in~\cite{Farina:2015uea}. In that work the twin QCD phase transition is raised by simply increasing the twin sector masses. However, the increase in mass required to reduce $\Delta N_\text{eff}$ introduces substantial tuning.

In the above analysis we have assumed that all of the hatted fields in the twin sector have masses above the decoupling scale. This is in some tension with the assumption of minimal flavor violation; see discussion in Sec.~\ref{sec:model}. Therefore, it may be that the reduction in $\Delta N_\text{eff}$ outlined above would be correlated with new signals of BSM flavor violation. If we strictly adhere to the MFV form, then there are new fermions in the twin sector bath at decoupling. One scenario would be to take $\lambda_{\widehat{U}\widehat{D}}=10\lambda_D$ and take $\cot\vartheta\approx 5$. In this case only two new Dirac fermions have masses below the decoupling scale and the $N_\text{eff}$ range becomes
\beq
N_\text{eff}\approx1.7\text{\textendash}2.0~.
\eeq

The ideas sketched above can be strengthened through quantitative estimates of the relevant parameters.  
The masses of mirror hadrons and other dynamical scales relevant to the mirror SU(4$)_c$ interaction can be estimated using simple scaling arguments.
First, the strong interaction in the mirror sector, with its larger beta function, confines at a higher scale than in the visible sector. Based on two-loop running, we find that 
\begin{equation}
\Lambda_{B} \approx (15 - 25) \Lambda_A ~,
\end{equation}
as shown in Fig.~\ref{fig:confinementScale} and detailed in Appendix~\ref{ap:Running}.

We first consider the temperature of the QCD phase transition in the twin sector. The chiral symmetry breaking phase transition temperature should be proportional to $f_\pi$, which we assume depends on $\Lambda$ like
\beq
f_\pi \propto \Lambda~.\label{eq:fPiScale}
\eeq
This estimate harmonizes with existing lattice results~\cite{Perez:2020vbn}. Therefore, we estimate
\begin{equation}
T_{c B} \approx  T_{c A} \frac{\Lambda_B}{\Lambda_A} \approx (15 - 25) T_{c A},
\end{equation}
where the SM transition temperature is $T_{c A} \approx  160$ MeV~\cite{HotQCD:2018pds}. The predicts a phase transition in the twin sector at about 2.5\textendash4 GeV. This quite close to (though slightly above)  the decoupling temperature $T_D\approx 2$ GeV of sectors, so order one numbers may impact how effectively the energy from the twin transition is transferred to the visible sector. Still, it is significant that the structure of this model, motivated by other consideration, acts to reduce the contribution to $\Delta N_\text{eff}$.

We would also like to know if the pions contribute to the degrees of freedom in the twin sector. The pion masses can be estimated from chiral perturbation theory. We use the Gell-Mann-Oakes-Renner relation~\cite{Gell-Mann:1968hlm}
\beq
m_\pi^2 \sim \frac{m_q}{f_\pi^2}\,  \langle 0 | q \bar q | 0\rangle~,
\eeq
where $m_q$ is the light quark mass, $\langle 0 | q \bar q | 0\rangle$ is the chiral-symmetry breaking condensate, and $f_\pi$ is the pion decay constant.  Neglecting the mild dependence on the number of colors, we assume the condensate scales as
\beq
\langle 0 | q \bar q | 0\rangle \propto \Lambda^3~.
\eeq
Using this and Eq.~\eqref{eq:fPiScale} the mirror pion masses may be estimated as 
\begin{equation}
m_{\pi B}^2 \approx m_{\pi A}^2 \, \cot \vartheta  \, \frac{\Lambda_B}{\Lambda_A} \gtrsim (45 - 75) m_{\pi A}^2~,
\end{equation}
or $m_{\pi B} \gtrsim (7 - 9)m_{\pi A}$. This puts the twin pion masses at about 1 GeV, which is just below the decoupling temperature of $T_D\approx2$ GeV. Therefore, they are likely to contribute to $g_{\ast B}$ and increase the value of $\Delta N_\text{eff}$ to be about one. This value is still in tension with experimental measurements, but is significantly lower than the standard twin Higgs value. This means that any other mechanism that might be added to the scenario to further reduce the contribution to $\Delta N_\text{eff}$ can be much less effective, opening up additional parameter regions.

Due to the unbroken SU(4$)_{c}$ symmetry, the baryons in the twin sector are bosonic states constructed of four spin-$\frac{1}{2}$ quarks. As in the visible sector, isospin is a good approximate symmetry of the theory, implying that the lightest baryon should be built of the lightest $u$ and $d$ quarks. The baryon wavefunction may be decomposed into factors describing the spatial, spin, color, and flavor degrees of freedom, and must be completely antisymmetric under exchange of any two quarks. The spatial wavefunction is completely symmetric for the lowest lying baryons with vanishing orbital angular momentum, while the color wavefunction is completely antisymmetric due to its color singlet nature. 
The remaining portion of the wavefunction, describing spin and flavor degrees of freedom, must therefore be completely symmetric. Under this condition, a straightforward isospin analysis suggests that the lightest baryon should be an isospin singlet, spin zero state composed of two up quarks and two down quarks. 

The generator of the unbroken twin electromagnetism after electroweak symmetry breaking is $Q = t^3 + X$, see Eq.~\eqref{eq:Echarge}. Thus, the electric charge of the lightest twin baryon is
\begin{equation}
Q = 2 \, Q_u + 2 \,Q_d = 2 \left( \frac{5}{8} \right) +  2\left( -\frac{3}{8} \right) = \frac{1}{2}~,
\end{equation}
where the $X$ charges of the quark fields are given in~\eqref{eq:Qcharges}.
Other quark combinations correspond to spin-1 and spin-2 baryons. From the results in~\cite{LSD:2014obp} the spin-1 baryon is about 3\% heavier than the spin-0 while the spin-2 baryon is about 11\% heavier. Differences in the constituent masses and charges affect these specific lattice predictions. Mass splittings of hundreds of MeV lead to prompt $\beta$-decay of the higher spin states into the spin-0 ground state. This implies in particular that only the lightest baryon is present on cosmological time scales. 

The lightest baryon masses in the SM can be approximately written as  $m_{p A} \approx 3 \hat m_{q A}$, where $\hat m_{q A}$ is the constituent quark masses. Assuming $\hat m_q \propto \Lambda$, we estimate the mass of the lightest twin baryon as 
\begin{equation}
m_{p B} \approx  4 \hat m_{q B}  \approx \, 4 \hat m_{q A}  \frac{\Lambda_B}{\Lambda_A} \approx  \frac{4}{3} \frac{\Lambda_B}{\Lambda_A} m_{p A} \approx (20 - 30) m_{p A}~.
\end{equation}
Including the contribution from the constituent quark spin-spin interaction yields a quantitatively similar estimate. Thus the lightest twin baryons have masses of a few tens of GeV. This, in turn, implies the next lightest baryon mass is more than half a GeV heavier than the spin-0 state and that $\beta$-decay occurs quickly.

If the cosmological population is composed entirely out of the spin-0 baryons then these would presumably combine with the twin electrons to create bound states. A single baryon with a single electron has electric charge of $-1/2$, leading to further electromagnetic interactions. The charge neutral bound states of twin baryons and twin electrons would seem to be something more like a molecule, with two baryons and one electron. The form is something like the covalent bond of $\text{H}^+$.

The presence of stable, electrically charged twin-sector states can induce substantial modifications to dark matter substructure on galactic scales through dissipative self-interactions, placing strong bounds on the fraction they contribute to the total dark matter density~\cite{Fan:2013yva,Ghalsasi:2017jna,Chang:2018bgx,Huo:2019yhk,Shen:2021frv,Shen:2022opd,Roy:2023zar,Gemmell:2023trd}. At the same time, atomic-like dark matter can have some cosmologically attractive features~\cite{Bansal:2022qbi}. Both these constraints, and these virtues, can be eliminated if the twin electric symmetry is broken. This can be done within the context of a mirror model by spontaneously breaking the twin hypercharge~\cite{Batell:2019ptb}. 

The cosmological implications of our scenario hinge on the radiation content in the mirror sector and the abundance of twin baryons, both of which are model-dependent. These factors can influence a range of phenomena, including mirror nucleosynthesis, recombination, twin baryon acoustic oscillations, and imprints on the CMB and large-scale structure. For recent studies of these issues in the minimal mirror twin Higgs framework, see Refs.~\cite{Chacko:2018vss,Bansal:2021dfh}. A comprehensive analysis lies beyond our present scope but represents an important direction for future work.

\section{Conclusion\label{sec:Con}}
In this work we describe a novel variation of the twin Higgs concept, in which the color group in both sectors is SU(4$)_c$ at high scales. This gauge symmetry is spontaneously broken to SU(3$)_c$ only in the visible sector, which spontaneously breaks $Z_2$ and allows for a viable electroweak vacuum alignment. This simple variation of the twin Higgs leads to several interesting consequences.

The color-breaking structure leads to several opportunities to discover this model at colliders. It produces a heavy neutral vector $Z'$ with much large branching into jets than leptons. This leads dilepton and dijet searches to have comparable sensitivity to this $Z'$, offering interesting opportunities to discover and study this state through multiple channels at the LHC.

Another prediction of this set-up is the existence of states with electric charge of $\pm1/2$, some of which have top-Yukawa strength coupling to the Higgs. The searches for fractionally charged particles at the LHC is still developing. This model provides a definite target for these searches and motivation to ensure the experimental results can be applied to a variety of theoretical scenarios.

While the BSM visible sector states associated with the top quark can have masses much lager than the top quark, we have shown that tuning associated with the Higgs potential is not severe. Indeed, by making a detailed analysis of the tuning in both the original mirror twin Higgs model as well as this extended color variation we show that the extended color tuning is often somewhat less that the standard twin Higgs. 

The unbroken SU(4$)_c$ structure of the twin sector also leads to interesting cosmological implications. The natural contributions to $\Delta N_\text{eff}$ are significantly less than the mirror twin Higgs. The lightest baryons of the twin sector are spin-0, leading to intriguing dark matter possibilities and signatures. Our present work has only begun the exploration of possible cosmological consequences. It would be interesting to further this analysis in future work.

\section*{Acknowledgments}
We thank Zackaria Chacko, Seth Koren, and Steven Lowette for useful communications. The work of B.B. is supported by the U.S. Department of Energy under grant No. DE–SC0007914.
T.C., L.P. and C.B.V acknowledge support from the National Science Foundation under Grant No. PHY-2210067.

\appendix

\section{SU(4) Generators\label{app:su4}}
We use the following representation of the SU(4) Lie Algebra 
\begin{align}
T^1=&\left(\begin{array}{cccc}
0 & 1 & 0 & 0 \\
1 & 0 & 0 & 0 \\
0 & 0 & 0 & 0 \\
0 & 0 & 0 & 0 
    \end{array} \right)~,& T^2=&\left(\begin{array}{cccc}
0 & -i & 0 & 0 \\
i & 0 & 0 & 0 \\
0 & 0 & 0 & 0 \\
0 & 0 & 0 & 0 
    \end{array} \right)~,& T^3=&\left(\begin{array}{cccc}
1 & 0 & 0 & 0 \\
0 & -1 & 0 & 0 \\
0 & 0 & 0 & 0 \\
0 & 0 & 0 & 0 
    \end{array} \right)~,&\nonumber \\
T^4=&\left(\begin{array}{cccc}
0 & 0 & 1 & 0 \\
0 & 0 & 0 & 0 \\
1 & 0 & 0 & 0 \\
0 & 0 & 0 & 0 
    \end{array} \right)~,& T^5=&\left(\begin{array}{cccc}
0 & 0 & -i & 0 \\
0 & 0 & 0 & 0 \\
i & 0 & 0 & 0 \\
0 & 0 & 0 & 0 
    \end{array} \right)~,& T^6=&\left(\begin{array}{cccc}
0 & 0 & 0 & 0 \\
0 & 0 & 1 & 0 \\
0 & 1 & 0 & 0 \\
0 & 0 & 0 & 0 
    \end{array} \right)~,&\nonumber \\
T^7=&\left(\begin{array}{cccc}
0 & 0 & 0 & 0 \\
0 & 0 & -i & 0 \\
0 & i & 0 & 0 \\
0 & 0 & 0 & 0 
    \end{array} \right)~,& T^8=&\frac{1}{\sqrt{3}}\left(\begin{array}{cccc}
1 & 0 & 0 & 0 \\
0 & 1 & 0 & 0 \\
0 & 0 & -2 & 0 \\
0 & 0 & 0 & 0 
    \end{array} \right)~,& T^9=&\left(\begin{array}{cccc}
0 & 0 & 0 & 1 \\
0 & 0 & 0 & 0 \\
0 & 0 & 0 & 0 \\
1 & 0 & 0 & 0 
    \end{array} \right)~,&\nonumber \\
T^{10}=&\left(\begin{array}{cccc}
0 & 0 & 0 & -i \\
0 & 0 & 0 & 0 \\
0 & 0 & 0 & 0 \\
i & 0 & 0 & 0 
    \end{array} \right)~,& T^{11}=&\left(\begin{array}{cccc}
0 & 0 & 0 & 0 \\
0 & 0 & 0 & 1 \\
0 & 0 & 0 & 0 \\
0 & 1 & 0 & 0 
    \end{array} \right)~,& T^{12}=&\left(\begin{array}{cccc}
0 & 0 & 0 & 0 \\
0 & 0 & 0 & -i \\
0 & 0 & 0 & 0 \\
0 & i & 0 & 0 
    \end{array} \right)~,&\nonumber \\
T^{13}=&\left(\begin{array}{cccc}
0 & 0 & 0 & 0 \\
0 & 0 & 0 & 0 \\
0 & 0 & 0 & 1 \\
0 & 0 & 1 & 0 
    \end{array} \right)~,& T^{14}=&\left(\begin{array}{cccc}
0 & 0 & 0 & 0 \\
0 & 0 & 0 & 0 \\
0 & 0 & 0 & -i \\
0 & 0 & i & 0 
    \end{array} \right)~,& T^{15}=&\frac{1}{\sqrt{6}}\left(\begin{array}{cccc}
1 & 0 & 0 & 0 \\
0 & 1 & 0 & 0 \\
0 & 0 & 1 & 0 \\
0 & 0 & 0 & -3 
    \end{array} \right)~,& 
\end{align}
which satisfy
\beq
\text{Tr}\left[ T^A T^B\right]=2\delta^{AB}~.
\eeq
The non-zero structure constants are
\begin{align}
f^{123}&=1~, \ \ f^{458}=f^{678}=\frac{\sqrt{3}}{2}~, \ \ f^{8\,9\,10}=f^{8\,11\,12}=\frac{1}{2\sqrt{3}}~, \ \ f^{8\,13\,14}=-\frac{1}{\sqrt{3}}  \nonumber\\
f^{147}&=f^{165}=f^{1\,9\,12}=f^{1\,11\,10}=f^{246}=f^{257}=f^{2\,9\,11}=f^{2\,10\,12}=f^{345}=f^{376}=f^{3\,9\,10}=\frac12~,\nonumber\\
f^{3\,12\,11}&=f^{4\,9\,14}=f^{4\,13\,10}=f^{5\,9\,13}=f^{5\,10\,14}=f^{6\,11\,15}=f^{6\,14\,12}=f^{7\,11\,13}=f^{7\,12\,14}=\frac12~,\nonumber\\
f^{9\,10\,15}&=f^{11\,12\,15}=f^{13\,14\,15}=\sqrt{\frac{2}{3}}
\end{align}

\section{BSM Vector and Fermion Masses and Couplings\label{a.Details}}
In this section we provide the details regarding new vector and fermion states that arise in this model. This includes their couplings and masses.

\subsection{BSM Vector Eigenstates\label{a.Vecs}}
In this section we provide the details regarding the BSM vectors in the visible sector. When $\Phi_A$,  which is in the fundamental representation of SU(4$)_{cA}$ and charged under U(1$)_{X\,A}$ gets a VEV of the form
\beq
\langle\Phi_A\rangle=\left( \begin{array}{c}
0 \\
0 \\
0 \\
w
\end{array}\right)~,
\eeq
the SU(3$)_c$ subgroup is preserved
\beq
\frac12 T^{1,\ldots,8}\langle\Phi_A\rangle=0~.
\eeq
We also note that the VEV is an eigenvector of one of the broken generators:
\beq
\frac12 T^{15}\langle\Phi_A\rangle=-\frac12\sqrt{\frac32}\langle\Phi_A\rangle~.
\eeq

We define the gauge boson of U(1$)_X$ as $X_\mu$ and the gauge coupling $g_X$. The gauge coupling of SU(4$)_c$ is denoted $g_s$ as it is inherited by the residual SU(3$)_c$ color group. The gauge bosons are denoted by $C^A_\mu$ where the superscript is the SU(4$)_c$ gauge index for. The complete gauge covariant derivative is given by
\beq
D_\mu\psi=\left(\partial_\mu-ig_XX_\psi X_\mu-i\frac{g}{2}t^aA^a_\mu-i\frac{g_s}{2}T^AC^A_\mu\right)\psi~,
\eeq
where $X_\psi$ is the U(1$)_X$ charge of $\psi$ and $g$ is the usual SU(2$)_L$ gauge coupling. The generators and gauge bosons of SU(2$)_L$ are $\frac12t^a$ and $A^a_\mu$, respectively.  

The SU(4$)_c$ gauge bosons decompose into the eight gluons of the SU(3$)_c$ subgroup, denoted $G^A_\mu$, as well as a vector
\begin{align}
    \xi_\mu&=\frac{1}{\sqrt{2}} \begin{pmatrix} C_\mu^9-i C_\mu^{10}\\ 
    C_\mu^{11}-i C_\mu^{12}\\ 
   C_\mu^{13}-i C_\mu^{14}\\ 
   \end{pmatrix}~,
\end{align}
which is in the fundamental representation of SU(3$)_c$. The field $C^{15}_\mu$ is an SU(3$)_c$ color singlet.

\subsection{Color Breaking}
When SU(4$)_c\times$U(1$)_X$ spontaneously breaks to SU(3$)_c\times$U(1$)_Y$, seven NGBs are produced. Six are eaten by $\xi_\mu$ and the seventh by a linear combination of $X_\mu$ and $C^{15}_\mu$. To determine the massive linear combination we consider the covariant derivative acting on $\langle\Phi_A\rangle$:
\beq
D_\mu\langle\Phi_A\rangle=-i\left(\begin{array}{c}
   \frac{g_s}{\sqrt{2}}\xi_\mu \\
   \frac{g_X}{8}  X_\mu-g_s\sqrt{\frac38}C^{15}_\mu
\end{array}\right)w~.
\eeq
This leads, from $|D_\mu\langle\Phi_A\rangle|^2$, to the mass terms
\beq
\frac{w^2g_s^2}{2}\xi^\dag_\mu\xi^\mu+w^2\left(\frac{g_X^2}{64}+\frac38g_s^2 \right)\left(C^{15\mu}c_X-X^\mu s_X\right)\left(C^{15\mu}c_X-X^\mu s_X \right)~,
\eeq
where
\beq
\cos\theta_X\equiv c_X=\frac{g_s \sqrt{24}}{\sqrt{24g_s^2+g_X^2}}~, \ \ \ \sin\theta_X\equiv s_X=\frac{g_X}{\sqrt{24g_s^2+g_X^2}}~. \label{eq:cosX}
\eeq
Therefore, we identify the masses
\beq
m_\xi\equiv \frac{g_sw}{\sqrt{2}}~, \ \ \ m_{B'}\equiv\frac{w}{4\sqrt{2}}\sqrt{24g_s^2+g_X^2}~,\label{eq:mXiMBprime}
\eeq
where we have defined the massive state
\beq
B'_\mu\equiv C_\mu^{15} c_X-X_\mu s_X~.
\eeq
The massless, orthogonal state is associated with SM hypercharge
\beq
B_\mu=X_\mu c_X+C_\mu^{15} s_X~.
\eeq
We invert this relationship and find
\begin{align}
    X_\mu=&\,B_\mu c_X-B'_\mu s_X~,& 
    C^{15}_\mu=&\,B'_\mu c_X+B_\mu s_X~,
\end{align}

For a generic field $\psi$ this means that the SU(4$)_c$ part of the covariant derivative can be written as
\beq
\frac{g_s}{2}T^AC^A_\mu=\frac{g_s}{2}\lambda^A G^A_\mu+\frac{g_s}{\sqrt{2}}\left(\begin{array}{cc}
   0  & \xi_\mu \\
   \xi^\dag_\mu  &0 
\end{array} \right)+g_s\sqrt{\frac38}T_c\left(B'_\mu c_X+B_\mu s_X \right)~,
\eeq
where the $\lambda^A$ matrices are the generators of SU(3$)_c$ and 
\beq
T_c=\left(\begin{array}{cccc}
    \frac13 & 0 & 0 & 0 \\
    0 & \frac13  & 0 & 0 \\
    0 & 0 & \frac13  & 0 \\
    0 & 0 & 0 & -1 
\end{array} \right)~.\label{eq:defTc}
\eeq
By including the U(1$)_X$ part we find 
\begin{align}
D_\mu\psi\supset&-iB'_\mu\left[g_s\sqrt{\frac38} c_X T_c-g_XX_\psi s_X\right]\psi
-ig'B_\mu\left[X_\psi+X_\Phi T_c \right]\psi~,
\end{align}
where the hypercharge gauge coupling is given by
\beq \label{g'gx}
g'\equiv\frac{g_Xg_s\sqrt{24}}{\sqrt{24g_s^2+g_X^2}}~.
\eeq
In short, we have normalized the U(1$)_X$ charges so that states in the $(\bm{4},X)$ representation of SU(4$)_c\times$U(1$)_X$ become 
\beq
(\bm{4},X)\to \left(\bm{3},X+\frac{1}{24}\right)\oplus\left(\bm{1},X-\frac18\right)
\eeq
under SU(3$)_c\times$U(1$)_Y$.
The relationship in \eqref{g'gx} can also be inverted, giving us the value of $g_X$ in terms of known quantities:
\beq
    g_X = \frac{\sqrt{24}g_s g'}{\sqrt{24 g_s^2 - g'^2}}~.
\eeq
From this we can rewrite the mixing parameters as
\beq
c_X=\sqrt{1-\frac{g^{\prime2}}{g_s^224}}~, \ \ \ s_X=\frac{g'}{g_s\sqrt{24}}~.
\eeq
This implies that at the scale of the SM $Z$ mass $s_X\approx0.06$ and $g_X\approx g'$.

\subsection{Electroweak Symmetry Breaking}
In the visible sector, when the Higgs gets a VEV
\beq
\langle H_A\rangle=\frac{1}{\sqrt{2}}\left(\begin{array}{c}
0\\
v_A
\end{array}\right)~,
\eeq
the usual electroweak symmetry breaking is modified because the Higgs ultimately couples to $X_\mu$ not just $B_\mu$. The $W^\pm_\mu$ are defined in the usual way
\beq
W^\pm_\mu=\frac{1}{\sqrt{2}}\left(A^1_\mu\mp i\, A^2_\mu \right)~,
\eeq
and receive the usual mass. However, the masses of the neutral bosons is more complicated. From $|D_\mu\langle H_A\rangle|^2$, and dropping the $W^\pm_\mu$ terms, we find
\beq
\frac{v_A^2}{8}\left(g_XX_\mu-gA_\mu^3\right)\left(g_XX^\mu-gA^{3\mu} \right)~.\label{e.Zbarmass}
\eeq
Using the usual definitions
\beq
\cos\theta_W\equiv c_W=\frac{g}{\sqrt{g^2+g^{\prime2}}}~, \ \ \sin\theta_W\equiv s_W=\frac{g'}{\sqrt{g^2+g^{\prime2}}}~,
\eeq
we define the massless photon and a massive boson
\begin{align}
    A_\mu=&\,B_\mu c_W+A^3_\mu s_W~,&
    \overline{Z}_\mu=&-B_\mu s_W+A^3_\mu c_W~.
\end{align}
Then the mass term in Eq.~\eqref{e.Zbarmass} becomes
\beq
\frac{v_A^2}{8}\left(B'_\mu g_Xs_X+\sqrt{g^2+g^{\prime2}}\overline{Z}_\mu\right)\left(B'_\mu g_Xs_X+\sqrt{g^2+g^{\prime2}}\overline{Z}^\mu\right)~.
\eeq
Thus we see that the Higgs gives an additional mass contribution to $B'_\mu$ and a mass mixing between $B'_\mu$ and $\overline{Z}_\mu$. The mass terms of the neutral bosons can be written as
\beq
\frac12\left(B'_\mu,\,\overline{Z}_\mu \right)\left(\begin{array}{cc}
 m_{B'}^2+\frac{v_A^2}{4}g_X^2s_X^2    & m_{Z_0}\frac{v_A}{2}g_Xs_X  \\
    m_{Z_0}\frac{v_A}{2}g_Xs_X & m^2_{Z_0}
\end{array}\right)\left(\begin{array}{c}
    B^{\prime\mu} \\
    \overline{Z}^\mu 
\end{array}\right)~,
\eeq
where 
\beq
m_{Z_0}=\frac{v_A}{2}\sqrt{g^2+g^{\prime2}}
\eeq 
is the SM $Z$ mass. The diagonal masses are
\begin{align}
&m^2_\pm=\\
&\frac12\left(m_{B'}^2+m_{Z_0}^2+\frac{v_A^2}{4}g_X^2s^2_X\pm\sqrt{(m_{B'}^2-m_{Z_0}^2)^2+(m_{B'}^2+m_{Z_0}^2)\frac{v_A^2}{2}g_X^2s^2_X+\frac{v_A^4}{16}g_X^4s^4_X} \right)~.\nonumber
\end{align}
The mass eigenstates are defined by
\begin{align}
    Z'_\mu=&\,B'_\mu c_M-\overline{Z}_\mu s_M~,&
    Z_\mu=&\,B'_\mu s_M+\overline{Z}_\mu c_M
\end{align}
where $m_+=m_{Z'}$ and $m_-=m_{Z}$. The mixing angles, $\sin\theta_M\equiv s_M$ and similar for the cosine, are defined by
\beq
\cos 2\theta_M=\frac{m_{B'}^2+\frac{v_A^2}{4}g_X^2s_X^2-m_{Z_0}^2}{m_{Z'}^2-m_{Z}^2}~, \ \ \sin 2\theta_M=-\frac{m_{Z_0}v_Ag_Xs_X}{m_{Z'}^2-m_{Z}^2}~.
\eeq
Because $v_A\ll w$ we can expand the mass eigenstates to find, to leading order in $v_A/w$, that
\beq \label{mZsq}
m_{Z'}^2\approx m_{B'}^2+\frac{v_A^2}{4}g_X^2s_X^2~, \ \ m_{Z}^2\approx m_{Z_0}^2-\frac{m_{Z_0}^2v_A^2g_X^2s_X^2}{4m_{B'}^2}~.
\eeq
Similarly, we find the mixing angle is about
\beq
\theta_M\sim-\frac{m_{Z_0}v_Ag^{\prime2}}{3w^2g_s^3\sqrt{6}}\approx -2\times 10^{-4}\left( \frac{1\text{ TeV}}{w}\right)^2~.
\eeq

We record the unbroken gauge fields in terms of the mass eigenstate fields 
\begin{align}
    X_\mu=&\,c_Xc_W A_\mu-(s_Xs_M+c_Xs_Wc_M)Z_\mu+(c_Xs_Ws_M-s_Xc_M)Z'_\mu\label{e.XinZZA} \\
    C^{15}_\mu=&\,s_Xc_W A_\mu+(c_Xs_M-s_Xs_Wc_M)Z_\mu+(s_Xs_Ws_M+c_Xc_M)Z'_\mu\label{e.C15inZZA} \\
    A^3_\mu=&\,s_W A_\mu+c_Wc_M Z_\mu-c_Ws_M Z'_\mu~.\label{e.A3inZZA}
\end{align}
This can also be inverted to find
\begin{align}
    A_\mu=&\,c_Wc_XX_\mu +s_W A^3_\mu+c_Ws_X C^{15}_\mu\\
    Z_\mu=&\,c_Mc_WA^3_\mu-(c_Ms_Wc_X+s_Ms_X)X_\mu+(s_Mc_X-c_Ms_Ws_X)C^{15}_\mu \\
    Z'_\mu=&\,(c_Mc_X+s_Ms_Ws_X)C^{15}_\mu-s_Mc_WA^3_\mu-(c_Ms_X+s_Ms_Wc_X)X_\mu~.
\end{align}
We see that the photon and $Z$ contain a small amount of $C_\mu^{15}$, meaning their couplings to quarks and leptons deviate only slightly from the SM values. The $Z'_\mu$ couples most strongly through $C_\mu^{15}$ giving it much larger couplings to SM quarks than to leptons. The couplings of these vectors to a field $\psi$ is given by 
\begin{align}
    &g_X X_\psi X_\mu+g t^3A^3_\mu+g_s\sqrt{\frac{3}{8}}T_cC^{15}_\mu=\nonumber\\
    &A_\mu\left[g_Xc_Xc_WX_\psi+gs_Wt^3+g_s\sqrt{\frac{3}{8}}s_Xc_WT_c \right]\nonumber\\
    &+Z_\mu\left[-g_X(s_Xs_M+c_Xs_Wc_M)X_\psi+gc_Wc_Mt^3+g_s\sqrt{\frac{3}{8}}(c_Xs_M-s_Xs_Wc_M)T_c \right]\nonumber\\
    &+Z'_\mu\left[g_X(c_Xs_Ws_M-s_Xc_M)X_\psi-gc_Ws_Mt^3+g_s\sqrt{\frac{3}{8}}(c_Xc_M+s_Xs_Ws_M)T_c \right]~.
\end{align}
We note that
\beq
e=gs_W=g'c_W~, \ \ \ g'=g_Xc_X=\sqrt{24}g_ss_X~.
\eeq
The final result is
\begin{align}
&g_X X_\psi X_\mu+g t^3A^3_\mu+g_s\sqrt{\frac{3}{8}}T_cC^{15}_\mu=\nonumber\\
&eQ_\psi A_\mu+Z_\mu\frac{g}{c_W}\left[c_M\left(t^3-s_W^2Q_\psi\right)+\frac{s_Ms_Wc_X}{8s_X}\left(T_c-8\frac{s_X^2}{c_X^2}X_\psi \right) \right]\nonumber\\
&+Z'_\mu\frac{g}{c_W}\left[\frac{c_Ms_Wc_X}{8s_X}\left( T_c-8\frac{s_X^2}{c_X^2}X_\psi \right)- s_M\left(t^3-s_W^2Q_\psi\right) \right]~.
\end{align}

\subsection{BSM Fermions\label{sa.bsmFerm}}
After SU(4$)_c$ color breaking in the visible sector we have the following mass terms amongst the BSM fermion field
\beq
\lambda_{\widehat{Q}}\omega\overline{Q}_4\widehat{Q}+\lambda_{\widehat{U}}\omega\overline{U}_4\widehat{U}+\lambda_{\widehat{D}}\omega\overline{D}_4\widehat{D}+\text{ H.c.}~,
\eeq
where the $\lambda_i$ couplings are taken to be proportional to the identity matrix in the appropriate flavor space. Electroweak symmetry breaking contributes to both
\beq
\frac{\lambda_Dv_A}{\sqrt{2}}\overline{D}_{L4}D_4+\frac{\lambda_Uv_A}{\sqrt{2}}\overline{U}_{L4}U_4+\text{ H.c.}~,
\eeq
and
\beq
\frac{\lambda_{\widehat{Q}\widehat{D}}v_A}{\sqrt{2}}\overline{\widehat{Q}}_D\widehat{D}+\frac{\lambda_{\widehat{Q}\widehat{U}}v_A}{\sqrt{2}}\overline{\widehat{Q}}_U\widehat{U}+\text{ H.c.}~.
\eeq
In this second term we again assume that the Yukawa matrices are proportional to the identity matrix in flavor space. These interactions imply that the fermion masses in each sector can be organized into the matrices 
\beq
\left(\overline{U}_4,\,\overline{\widehat{Q}}_U \right)\mathcal{M}_U\left(\begin{array}{c}
    \widehat{U} \\
    U_{L4}
\end{array} \right)~, \ \ \ \left(\overline{D}_4,\,\overline{\widehat{Q}}_D \right)\mathcal{M}_D\left(\begin{array}{c}
    \widehat{D} \\
    D_{L4}
\end{array} \right)~,
\eeq
with
\beq
\mathcal{M}^A_U=\left(\begin{array}{cc}
    \lambda_{\widehat{U}}w & \lambda_{U}^\dag\frac{v_A}{\sqrt{2}} \\
    \lambda_{\widehat{Q}\widehat{U}}\frac{v_A}{\sqrt{2}} & \lambda_{\widehat{Q}}w
\end{array} \right)~, \ \ \ \mathcal{M}^A_D=\left(\begin{array}{cc}
    \lambda_{\widehat{D}}w & \lambda_{D}^\dag\frac{v_A}{\sqrt{2}} \\
    \lambda_{\widehat{Q}\widehat{D}}\frac{v_A}{\sqrt{2}} & \lambda_{\widehat{Q}}w
\end{array} \right)~,
\eeq
and
\beq
\mathcal{M}^B_U=\left(\begin{array}{cc}
    0 & \lambda_{U}^\dag\frac{v_B}{\sqrt{2}} \\
    \lambda_{\widehat{Q}\widehat{U}}\frac{v_B}{\sqrt{2}} & 0
\end{array} \right)~, \ \ \ \mathcal{M}^B_D=\left(\begin{array}{cc}
    0 & \lambda_{D}^\dag\frac{v_B}{\sqrt{2}} \\
    \lambda_{\widehat{Q}\widehat{D}}\frac{v_B}{\sqrt{2}} & 0
\end{array} \right)~,
\eeq

The $B$ sector mass are already diagonal, but in the $A$ sector there is mixing. In most cases we neglect the mixing because the Yukawa matrices multiplied by the electroweak VEV are small, no larger than about the bottom-quark Yukawa coupling. Of course, the top quark is an exception. 
 The top-type masses are
\begin{align}
&m_{T\pm}^2=\frac12\left[w^2(\lambda_{\widehat{U}}^2+\lambda_{\widehat{Q}}^2)+m_t^2\pm\sqrt{\left(\lambda_{\widehat{U}}^2w^2+\lambda_{\widehat{Q}}^2w^2+m_t^2 \right)^2-4\lambda_{\widehat{U}}^2\lambda_{\widehat{Q}}^2w^4}~ \right]~,
\end{align}
where we have neglected smaller corrections from $\lambda_{\widehat{Q}\widehat{U}}$. To leading order in $m_t^2/w^2$ these are
\beq
m_{T+}^2=w^2\lambda_{\widehat{U}}^2+\mathcal{O}\left(\frac{m_t^2}{w^2} \right)~, \ \ m_{T-}^2=w^2\lambda_{\widehat{Q}}^2+\mathcal{O}\left(\frac{m_t^2}{w^2} \right)~.\label{eq:fermMass}
\eeq
This shows that these fermions have $w$-scale masses, as long as the $\lambda_{\widehat{U},\widehat{Q}}$ are not too small. The physical states are given by 
\begin{align}
    T_{L+}=&\, c_{TL}\widehat{U}-s_{TL}U_{L4}~,&  T_{L-}=&\, s_{TL}\widehat{U}+c_{TL}U_{L4}~, \\
    T_{R+}=&\, c_{TR}U_4-s_{TR}\widehat{Q}_U~,&  T_{R-}=&\,s_{TR}U_4+c_{TR}\widehat{Q}_U~,
\end{align}
where 
\begin{align}
    \cos2\theta_{TL}&=\frac{w^2(\lambda_{\widehat{U}}^2-\lambda_{\widehat{Q}}^2)-m_t^2}{m_{T+}^2-m_{T-}^2}~,& \sin2\theta_{TL}&=-2\frac{m_tw\lambda_{\widehat{U}}}{m_{T+}^2-m_{T-}^2}~,\\
    \cos2\theta_{TR}&=\frac{w^2(\lambda_{\widehat{U}}^2-\lambda_{\widehat{Q}}^2)+m_t^2}{m_{T+}^2-m_{T-}^2}~,& \sin2\theta_{TR}&=-2\frac{m_tw\lambda_{\widehat{Q}}}{m_{T+}^2-m_{T-}^2}~.
\end{align}
Neglecting the mixing in all cases but the top flavored case, we write the magnitude of the Yukawas in terms of the physical masses
\begin{align}
\lambda_{\widehat{U}}^2=&\,\frac12\left[\frac{m_{T+}^2+m_{T-}^2-m_t^2}{w^2}+\sqrt{\left(\frac{m_{T+}^2+m_{T-}^2-m_t^2}{w^2} \right)^2-4\frac{m_{T+}^2m_{T-}^2}{w^4}} \right]~,\\
\lambda_{\widehat{Q}}^2=&\,\frac12\left[\frac{m_{T+}^2+m_{T-}^2-m_t^2}{w^2}-\sqrt{\left(\frac{m_{T+}^2+m_{T-}^2-m_t^2}{w^2} \right)^2-4\frac{m_{T+}^2m_{T-}^2}{w^4}} \right]~.
\end{align}
Note that in order for these magnitudes to be real we must have that the mass states are split by at least the top quark mass. 

\subsection{Running Strong Coupling\label{ap:Running}}
In this work we use two-loop running of the strong couplings $\alpha_{sA}$ and $\alpha_{sB}$ in the visible and twin sectors, respectively. In each sector we use the formula given in~\cite{Ellis:1996mzs}
\beq
b_0\ln\frac{Q^2}{Q_0^2}=\frac{1}{\alpha_s(Q^2)}-\frac{1}{\alpha_s(Q_0^2)}+b_1\ln\frac{\alpha_s(Q^2)\left[1+b_1\alpha_s(Q_0^2) \right]}{\alpha_s(Q_0^2)\left[1+b_1\alpha_s(Q^2) \right]}~,
\eeq
where $Q$ and $Q_0$ are the energy scales being compared. For SU($N$) gauge theories with $n_f$ fermions in the fundamental representation
\beq
b_0=\frac{11N-2n_f}{12\pi}~, \ \ \ \ b_1=\frac{17N^2-5Nn_f-\frac{3n_f}{2N}(N^2-1)}{2\pi(11N-2n_f)}~.
\eeq

We note that these formulae do not account for the massive $\xi^\mu$ vector which is a color fundamental. When this vector is included in the SU(3$)_c$ running is essentially running like SU(4$)_c$ because the gluons that were ``lost" by being too massive have been restored. This means that we expect the visible and twin couplings to run identically from the color-breaking scale $w$ down to $m_\xi$. 

So, to determine the running of $\alpha_{sB}$ we use $\alpha_{sA}(m_\xi)=\alpha_{sB}(m_\xi)$ as our starting point for running to lower energy scales. We can then use $\alpha_{sB}$ to estimate the twin confinement scale. Following Ref.~\cite{Ellis:1996mzs} we use the formula
\beq
    \Lambda_B=Q\exp\left[-\frac{1}{2b_0\alpha_s(Q)}\right]\left(\frac{b_1}{b_0}+\frac{1}{b_0\alpha_s(Q)} \right)^{b_1/(2b_0)}~,
\eeq
to define the confining scale $\Lambda_B$ of the twin sector. 

\section{Useful Phenomenology Formulae\label{a.Pheno}}
In this appendix we record useful formulae for determining the phenomenology of the $Z'$ boson. For any massive vector $V_\mu$ with couplings to a fermion $f$
\beq
\overline{f}\gamma^\mu\left(g_LP_L+g_RP_R \right)fV_\mu~,
\eeq
where 
\beq
P_{L,R}=\frac{1}{2}\left(1\mp\gamma_5 \right)~,
\eeq
the decay width into fermions of mass $m_f$ is given by
\beq
\Gamma(V\to ff)=\frac{N_c m_V}{24\pi }\sqrt{1-\frac{4m_f^2}{m_V^2}}\left[\left(g_L^2+g_R^2 \right)\left(1-\frac{m_f^2}{m_V^2} \right)+6g_Lg_R\frac{m_f^2}{m_V^2} \right]~.
\eeq
Decays to colored states are multiplied by a factor of
\beq
1+\frac{\alpha_s(m_V)}{\pi}~,
\eeq
to account for the leading QCD corrections. 

The decay width of $Z'$ into $W^+W^-$ is given by
\beq
\Gamma(Z'\to WW)=g_{Z'WW}^2\frac{m_{Z'}}{192\pi}\left(1-\frac{4m_W^2}{m_{Z'}^2} \right)^{3/2}\left(12+20\frac{m_{Z'}^2}{m_W^2}+\frac{m_{Z'}^4}{m_W^4} \right)~,
\eeq
while the width into $Zh$ is 
\beq
\Gamma(Z'\to Zh)=\frac{g^2_{hZZ'}}{192\pi}\frac{m_{Z'}}{m_{Z}^2}\lambda\left(\frac{m_Z^2}{m_{Z'}^2},\,\frac{m_h^2}{m_{Z'}^2} \right)\left[\lambda\left(\frac{m_Z^2}{m_{Z'}^2},\,\frac{m_h^2}{m_{Z'}^2} \right)^2+12\frac{m_Z^2}{m_{Z'}^2} \right]~,
\eeq
where
\beq
\lambda(a,b)=\sqrt{1-2(a+b)+(a-b)^2}~.
\eeq

The partonic production cross section of the $Z'$ through Drell-Yan ($\overline{q}q$) into a $\overline{f}f$ final state is given by~\cite{Ellis:1996mzs}
\beq
\sigma(s)=\frac{s N_c^f\left(g^2_{Li}+g^2_{Ri} \right)\left(g^2_{Lf}+g^2_{Rf} \right)}{48\pi N_c^i\left[(s-m_{Z'}^2)^2+\Gamma_{Z'}^2m_{Z'}^2 \right]}~,
\eeq
where $N_c^i=3$ is the usual SM quark colors from production through the initial particles and $N_c^f$ is the number of colors of the final state fermions. Because $\Gamma_{Z'}/m_{Z'}$ can be nearly 10\% in some cases, the narrow width approximation is not used in our analyses. 

For fermion production through all neutral gauge bosons, with couplings
\beq
eQ_f\overline{f}\gamma^\mu f+\overline{f}\gamma^\mu\left(g_{Lf}P_L+g_{Rf}P_R \right)fZ_\mu+\overline{f}\gamma^\mu\left(g'_{Lf}P_L+g'_{Rf}P_R \right)fZ'_\mu
\eeq
the cross section is
\begin{align}
\sigma(s)=\frac{N_c^f}{192\pi N_c^i}\sqrt{1-\frac{4m_f^2}{s}}&\left[\left( |C_{vv}|^2+|C_{av}|^2\right)(s+2m_f^2)\right.\nonumber\\
&\left.+\left( |C_{va}|^2+|C_{aa}|^2\right)(s-4m_f^2)\right]~,
\end{align}
where
\begin{align}
    C_{vv}=&\,\frac{4e^2Q_iQ_f}{s}+\frac{(g_{Ri}+g_{Li})(g_{Rf}+g_{Lf})}{s-m_Z^2}+\frac{(g'_{Ri}+g'_{Li})(g'_{Rf}+g'_{Lf})}{s-m_{Z'}^2+im_{Z'}\Gamma_{Z'}}~,\\
    C_{va}=&\,\frac{(g_{Ri}+g_{Li})(g_{Rf}-g_{Lf})}{s-m_Z^2}+\frac{(g'_{Ri}+g'_{Li})(g'_{Rf}-g'_{Lf})}{s-m_{Z'}^2+im_{Z'}\Gamma_{Z'}}~,\\
    C_{av}=&\,\frac{(g_{Ri}-g_{Li})(g_{Rf}+g_{Lf})}{s-m_Z^2}+\frac{(g'_{Ri}-g'_{Li})(g'_{Rf}+g'_{Lf})}{s-m_{Z'}^2+im_{Z'}\Gamma_{Z'}}~,\\
    C_{aa}=&\,\frac{(g_{Ri}-g_{Li})(g_{Rf}-g_{Lf})}{s-m_Z^2}+\frac{(g'_{Ri}-g'_{Li})(g'_{Rf}-g'_{Lf})}{s-m_{Z'}^2+im_{Z'}\Gamma_{Z'}}~.
\end{align}

The heavy BSM fermions can also be produced through $t$-channel $\xi^\mu$ exchange. At leading order this is captured by a single diagram of an incoming quark and antiquark (a up and antidown for example) with momenta $p_f$ and $p_{\overline{f}}$ producing a BSM fermion ($q_F$) and antifermion ($q_{\overline{F}}$).

In the center of momentum frame, and neglecting the masses of the initial state quarks we find
\begin{align}
&\frac{d\sigma}{d\Omega}=\frac{g_s^4\beta}{N_c64\pi^2s}\frac{4}{(s+2m_\xi^2-m_F^2-m_{\overline{F}}^2-s\beta\cos\theta)^2}\frac{s^2}{16}\\
&\left\{2\left[5-4\frac{m_F^2+m_{\overline{F}}^2}{s}-\frac{(m_F^2-m_{\overline{F}}^2)^2}{s^2} \right]+\frac{m_F^2m_{\overline{F}}^2}{m_\xi^2}\left[ 1-16\frac{m_\xi^2}{s}-\frac{(m_F^2-m_{\overline{F}}^2)^2}{s^2}\right]\right.\nonumber\\
&\left.+2\beta\cos\theta\left(2-\frac{m_F^2m_{\overline{F}}^2}{m_\xi^4} \right)+\beta^2\cos^2\theta\left(2+\frac{m_F^2m_{\overline{F}}^2}{m_\xi^4} \right)\right\}~.
\end{align}
The total cross section is
\begin{align}
    \sigma=&\,\frac{\alpha_s^2\pi}{2N_cs}\left\{\frac{\beta}{m_\xi^4+m_\xi^2(s-m_F^2-m_{\overline{F}}^2)+m_F^2m_{\overline{F}}^2}\left[(m_\xi^2-m_F^2)(m_\xi^2-m_{\overline{F}}^2)\left(2+\frac{m_F^2m_{\overline{F}}^2}{m_\xi^4} \right)\right.\right.\nonumber\\
&\left.+3sm_\xi^2\left(1-\frac{m_F^2m_{\overline{F}}^2}{2m_\xi^4}-\frac23\frac{m_F^2+m_{\overline{F}}^2}{m_\xi^2}   \right)+2s^2 \right]\nonumber\\
&\left.-4\left[1+\frac{1}{s}\left(1-\frac{m_F^2+m_{\overline{F}}^2}{2m_\xi^2}  \right)\left(1+\frac{m_F^2m_{\overline{F}}^2}{2m_\xi^4}  \right) \right]\tanh^{-1}\frac{\beta}{1+(2m_\xi^2-m_F^2-m_{\overline{F}}^2)/s}\right\}~.
\end{align}

\section{Mirror Twin Higgs Tuning\label{ap:tunig}}
In this appendix we discuss the twin Higgs potentials and characterizations of tuning. We begin with standard twin Higgs scenario and then consider the extended color model that is the focus of this work.

In the mirror twin Higgs framework the scalar potential is 
\beq
V=-M_H^2|\mathcal{H}|^2+\lambda_H|\mathcal{H}|^4+\delta_H\left(|H_A|^4+|H_B|^4 \right)+m^2\left(|H_A|^2-|H_B|^2\right)~,
\eeq
with 
\beq
\mathcal{H}=\left(\begin{array}{c}
H_A\\
H_B
\end{array}\right)
\eeq
where each of $H_A$ and $H_B$ are complex doublet fields. We parameterize them nonlinearly as
\beq
H_A=\bm{h}\frac{f}{\sqrt{\bm{h}^\dag\bm{h}}}\sin\left( \frac{\sqrt{\bm{h}^\dag\bm{h}}}{f}\right)~, \ \ \ \ H_B=\left(\begin{array}{c}
    0  \\
    \displaystyle f\cos\left( \frac{\sqrt{\bm{h}^\dag\bm{h}}}{f}\right)
\end{array} \right)~,
\eeq
where the field $\bm{h}$ is a doublet that obtains the VEV
\beq
\langle\bm{h}\rangle=\frac{1}{\sqrt{2}}\left(
\begin{array}{c}
0\\
v
\end{array}\right)~.
\eeq
Note that the VEVs of $H_A$ and $H_B$ are
\beq
\langle H_A\rangle=\left( \begin{array}{c}
0\\
f\sin\frac{v}{f\sqrt{2}}
\end{array}\right) ~, \ \ \ \ \langle H_B\rangle=\left( \begin{array}{c}
0\\
f\cos\frac{v}{f\sqrt{2}}
\end{array}\right) ~,
\eeq
This motivates the notation $\vartheta\equiv v/(f\sqrt{2})$.

We find in this and what follows that is helps to write the scalar potential as
\beq
V\left(\sqrt{\bm{h}^\dag\bm{h}}\right)~.
\eeq
We then define the vacuum of the potential by
\beq
\left.\frac{\partial V}{\partial\bm{h}}\right|_{\bm{h}\to\langle\bm{h}\rangle}=0~.
\eeq
We note that
\beq
\frac{\partial V}{\partial\bm{h}}=V'\left(\sqrt{\bm{h}^\dag\bm{h}}\right)\frac12\sqrt{\frac{\bm{h}^\dag}{\bm{h}}}~,
\eeq
where
\beq
V'=\frac{dV(x)}{dx}~.
\eeq
This shows that the vacuum is given by
\beq
V'\left(v/\sqrt{2}\right)=0~.\label{e.vacDef}
\eeq

The $\mu^2$ term is defined by
\beq
\mu^2=-\left.\frac{\partial^2V}{\partial\bm{h}^\dag\partial\bm{h}}\right|_{\bm{h}\to0}~,
\eeq
where
\beq
\frac{\partial^2V}{\partial\bm{h}^\dag\partial\bm{h}}=\frac14V''\left(\sqrt{\bm{h}^\dag\bm{h}}\right)+\frac{V'\left(\sqrt{\bm{h}^\dag\bm{h}}\right)}{4\sqrt{\bm{h}^\dag\bm{h}}}~.
\eeq
In other words,
\beq
\mu^2=-\frac14 V''(0)-\frac14\lim_{x\to0}\frac{V'(x)}{x}~.\label{e.muDef}
\eeq
Similarly, the mass of the Higgs boson is given by
\beq
m_h^2=\left.\frac{\partial^2\;}{\partial h^2}V\left(\frac{v+h}{\sqrt{2}}\right)\right|_{h\to0}~.
\eeq
This can be expressed as
\beq
m_h^2=\frac12V''(v/\sqrt{2})~.\label{e.mhDef}
\eeq

Dropping constant terms we find the potential can be written as
\begin{align}
V(x)=&\,-\frac{\delta_H f^4}{2}\sin^2\left( \frac{2x}{f}\right)-m^2f^2\cos\left( \frac{2x}{f}\right)~.
\end{align}
This quickly leads to 
\begin{align}
    V'(x)=&\,2f\sin\left( \frac{2x}{f}\right)\left[m^2-f^2\delta_H\cos\left( \frac{2x}{f}\right) \right]~,\\
    V''(x)=&\,4m^2\cos\left( \frac{2x}{f}\right)-4f^2\delta_H\cos\left( \frac{4x}{f}\right)~.
\end{align}
To determine the condition for stable vacuum we use Eq.~\eqref{e.vacDef} to find 
\beq
m^2=\delta_H f^2\cos(2\vartheta)~.\label{e.vacCon}
\eeq
Note that if $m=0$ that $\vartheta=\pi/4$ and so $H_A$ and $H_B$ have equal VEVs. 

We note that
\beq
\lim_{x\to0}\frac{V'(x)}{x}=4\left(m^2-f^2\delta_H \right)~.
\eeq
The Higgs mass parameter is obtained from Eq.~\eqref{e.muDef}, yielding
\begin{align}
\mu^2=&-\frac14 4\left(m^2-f^2\delta_H \right)-\frac144\left(m^2-f^2\delta_H \right)=2\left(f^2\delta_H -m^2 \right) ~.\label{e.muTH}
\end{align}
From Eq.~\eqref{e.mhDef} we find the Higgs mass to be
\begin{align}
m_h^2=&\,2m^2\cos(2\vartheta)-2f^2\delta_H\cos(4\vartheta)\nonumber\\
=&\,2\delta_Hf^2\sin^22\vartheta~,
\end{align}
where we have used the vacuum constraint in Eq.~\eqref{e.vacCon} in the second line. For this to be positive, and hence be a local minimum, we require that $\delta_H>0$.

We can use the vacuum constraint again to write
\begin{align}
m_h^2=&\,2f^2\delta_H\left(1-\frac{m^4}{\delta_H^2f^4} \right)~.
\end{align}
This last form makes clear that if $m=0$ that the Higgs mass is $\sqrt{2\delta_H}f$ and that a nonzero $m$ reduces the mass. We also see, from Eq.~\eqref{e.muTH}, that 
\beq
m_h^2=2\mu^2\cos^2\vartheta~.
\eeq
When $\vartheta\to0$ this gives the usual relation of $m_h^2=2\mu^2$.

We can now calculate the Barbieri-Giudice tuning of the $\mu^2$ parameter. For a parameter $x$, this is given by the absolute value of 
\beq
\Delta_x\equiv\frac{x}{\mu^2}\frac{\partial \mu^2}{\partial x}~.
\eeq
In this formula we first evaluate the derivative and only afterward use the vacuum constraint to simplify. We find for UV parameters $m^2$, $\delta_H$, and $f^2$ that 
\begin{align}
\Delta_{f^2}=&\,\Delta_{\delta_H}=\frac{1}{2\sin^2\vartheta}\approx\frac{f^2}{v^2}~,& \Delta_{m^2}=&\,1-\Delta_{f^2}\approx1-\frac{f^2}{v^2}~.
\end{align}
The tuning with respect to each of these three parameters is qualitatively the same. 

\subsection{Loop Corrections}
One-loop corrections to the tree-level potential can be obtained from the Coleman-Weinberg~\cite{Coleman:1973jx} potential
\beq
V_\text{CW}=-\frac{N_c}{8\pi^2}\Lambda^2\text{Tr}
\mathcal{M}^2-\frac{N_c}{16\pi^2}\text{Tr}\left[\mathcal{M}^4\left(\ln\frac{\mathcal{M}^2}{\Lambda^2}-\frac12 \right) \right]~,
\eeq
where $\Lambda$ is the energy cutoff of the theory. We evaluate this for fermions that couple to the Higgs. The fermion mass matrix in the top quark sector provides the largest effects and is given by 
\beq
\left(t_{RA},\,t_{RB} \right)\left(\begin{array}{cc}
   \lambda_t H_A   & 0 \\
    0 & \lambda_t H_B   
\end{array} \right)\left(\begin{array}{c}
    Q_{A} \\
    Q_{B}
\end{array} \right)~.
\eeq
The $\Lambda^2$ term of the potential is independent of $h$ because
\beq
\text{Tr}\mathcal{M}^\dag\mathcal{M} =\lambda_t^2f^2\text{Tr}\left(\begin{array}{cc}
   \sin^2\left( \frac{\sqrt{\bm{h}^\dag\bm{h}}}{f}\right)  & 0 \\
   0  & \cos^2\left( \frac{\sqrt{\bm{h}^\dag\bm{h}}}{f}\right)
\end{array} \right)=\lambda_t^2f^2=m_{tA}^2+m_{tB}^2~.
\eeq
 We next evaluate
\begin{align}
&\text{Tr}\left[\mathcal{M}^4\left(\ln\frac{\mathcal{M}^2}{\Lambda^2}-\frac12 \right) \right]=\\
&\lambda_t^4f^4\left\{\sin^4\left(\frac{\sqrt{\bm{h}^\dag\bm{h}}}{f} \right) \left[\ln\frac{\lambda_t^2f^2\sin^2\left(\frac{\sqrt{\bm{h}^\dag\bm{h}}}{f} \right) }{\Lambda^2}-\frac12 \right]+\cos^4\left(\frac{\sqrt{\bm{h}^\dag\bm{h}}}{f} \right) \left[\ln\frac{\lambda_t^2f^2\cos^2\left(\frac{\sqrt{\bm{h}^\dag\bm{h}}}{f} \right) }{\Lambda^2}-\frac12 \right] \right\}\nonumber
\end{align}
In other words, after dropping constant terms, we have
\begin{align}
&V_\text{CW}(x)=\\
&-\frac{3\lambda_t^4f^4}{16\pi^2}\left[\sin^4\left( \frac{x}{f}\right)\left(\ln\frac{\lambda_t^2f^2\sin^2\left( \frac{x}{f}\right)}{\Lambda^2} -\frac12\right)+\cos^4\left( \frac{x}{f}\right)\left(\ln\frac{\lambda_t^2f^2\cos^2\left( \frac{x}{f}\right)}{\Lambda^2} -\frac12\right) \right]~.\nonumber
\end{align}

In general, the $\Lambda^2$ part of the potential is independent of $x$. We can drop it and write
\beq
V_\text{CW}=-\frac{1}{16\pi^2}\sum_iN_{c_i}\left[\left(m_i^2\right)^2\ln\frac{m_i^2}{\Lambda^2}-\frac12\left(m_i^2\right)^2\right]~,\label{e.CWpot}
\eeq
where each squared mass is a function of $x$. It is then a straightforward exercise to determine
\beq
\frac{dV_\text{CW}}{dx}=-\frac{1}{8\pi^2}\sum_iN_{c_i}\frac{dm_i^2}{dx}m_i^2\ln\frac{m_i^2}{\Lambda^2}~,
\eeq
and
\beq
\frac{d^2V_\text{CW}}{dx^2}=-\frac{1}{8\pi^2}\sum_iN_{c_i}\left[ \left( \frac{dm_i^2}{dx}\right)^2\left(1+\ln\frac{m_i^2}{\Lambda^2} \right) + \frac{d^2m_i^2}{dx^2}m_i^2\ln\frac{m_i^2}{\Lambda^2}\right]~.
\eeq

For the present case this leads to
\begin{align}
    V_\text{CW}'(x)=&-\frac{3\lambda_t^4f^3}{8\pi^2}\sin\frac{2x}{f}\left(\sin^2\frac{x}{f}\ln\frac{\lambda_t^2f^2\sin^2\frac{x}{f}}{\Lambda^2}-\cos^2\frac{x}{f}\ln\frac{\lambda_t^2f^2\cos^2\frac{x}{f}}{\Lambda^2} \right) \nonumber\\
    =&\,\frac{3\lambda_t^4f^3}{8\pi^2}\sin\frac{2x}{f}\left( \cos\frac{2x}{f}\ln\frac{\lambda_t^2f^2\sin\frac{2x}{f}}{2\Lambda^2}+\ln\cot\frac{x}{f}\right)~,\\
    V_\text{CW}''(x)=&\,\frac{3\lambda_t^4f^2}{4\pi^2}\left( \cos\frac{4x}{f}\ln\frac{\lambda_t^2f^2\sin\frac{2x}{f}}{2\Lambda^2}+\cos\frac{2x}{f}\ln\cot\frac{x}{f}-\sin^2\frac{2x}{f}\right)
\end{align}

How does this affect the vacuum of the theory? We find
\begin{align}
V_\text{CW}'(v/\sqrt{2})=&\,2f\sin(2\vartheta)\frac{3\lambda_t^4}{16\pi^2}\left( \cos2\vartheta\ln\frac{\lambda_t^2f^2\sin2\vartheta}{2\Lambda^2} +\ln\cot\vartheta\right)
\end{align}
The full vacuum constraint is therefore modified to
\begin{align}
m^2=\delta_Hf^2\cos2\vartheta-\frac{3\lambda_t^4f^2}{16\pi^2}\left[\cos2\vartheta\ln\frac{\lambda_t^2f^2\sin2\vartheta}{2\Lambda^2}+\ln\cot\vartheta \right]~.
\end{align}
Note that the $m^2=0$ limit still implies $\vartheta=\pi/4$. 

The contribution to $\mu^2$ is
\begin{align}
\mu^2_\text{Loop}=&-\frac{3\lambda_t^4f^2}{8\pi^2}\ln\frac{\lambda_t^2f^2}{\Lambda^2}~.
\end{align}
The total mass parameter is then
\begin{align}
\mu^2=&\,2\left(\delta_Hf^2-m^2 \right)+\frac{3\lambda_t^4f^2}{8\pi^2}\ln\frac{\Lambda^2}{\lambda_t^2f^2}~.
\end{align}
Enforcing the constraint on $m^2$ we can write this as
\beq
\mu^2=2\sin^2\vartheta\left(2\delta_Hf^2-\frac{3f^2\lambda_t^4}{8\pi^2}\ln\frac{\lambda_t^2f^2\sin2\vartheta}{2\Lambda^2} \right)+\frac{3f^2\lambda_t^4}{8\pi^2}\ln\cos^2\vartheta
\eeq

The loop contribution to the Higgs mass is
\beq
m_{h\,\text{Loop}}^2=\frac{3\lambda_t^4f^2}{8\pi^2}\left(\cos4\vartheta\ln\frac{\lambda_t^2f^2\sin2\vartheta}{2\Lambda^2}+\cos2\vartheta\ln\cot\vartheta-\sin^22\vartheta \right)~.
\eeq
The total Higgs mass, after using the vacuum condition, is
\beq
m_h^2=2f^2\sin^22\vartheta\left[\delta_H-\frac{3\lambda_t^4}{16\pi^2}\left(1+\ln\frac{\lambda_t^2f^2\sin2\vartheta}{2\Lambda^2} \right)\right] ~,
\eeq
which means that $\mu^2$ and $m_h^2$ are related according to
\beq
m_h^2=2\cos^2\vartheta\mu^2-\frac{3\lambda_t^4f^2}{8\pi^2}\left[\sin^22\vartheta+2\cos^2\vartheta\ln\cos^2\vartheta\right]~.
\eeq
We can also relate the quartic coupling to the Higgs mass via
\beq
\delta_H=\frac{m_h^2}{2f^2\sin^2(2\vartheta)}+\frac{3\lambda_t^4}{16\pi^2}\left(1+\ln \frac{f^2\lambda_t^2\sin(2\vartheta)}{2\Lambda^2} \right)~,
\eeq
which means that
\beq
\mu^2=\frac{m_h^2}{2\cos^2\vartheta}+\frac{3\lambda_t^4f^2}{4\pi^2}\left(\sin^2\vartheta+\ln\cos\vartheta \right)~.
\eeq
Note that as $\vartheta\to0$ that $\mu^2\to m_h^2/2$. Also, we can write the vacuum condition as
\beq
m^2=\frac{m_h^2\cos2\vartheta}{2\sin^22\vartheta}+\frac{3\lambda_t^4f^2}{16\pi^2}\left(\cos2\vartheta+\ln\tan\vartheta \right)~.
\eeq

To calculate the tuning we take the appropriate derivatives before including the constraint from the vacuum. We then impose the vacuum constraint and expand in $v/f$. It is useful to define the quantity
\beq
\varpi=\frac{3\lambda_t^4f^2\sin^22\vartheta}{8\pi^2m_h^2}~.
\eeq
The tuning is then
\begin{align}
\Delta_{m^2}=&-\frac12\frac{\cos2\vartheta+\frac{\varpi}{1+\varpi}\ln\tan\vartheta}{\sin^2\vartheta+\frac{\varpi}{1+\varpi}\ln\cos\vartheta}=-\frac{f^2}{v^2}+\frac{3\lambda_t^4f^2}{4\pi^2m_h^2}\ln\frac{f}{v}+\ldots \label{eq:THmTune}\\
\Delta_{\delta_H}=&\,\frac{1+\frac{\varpi}{1+\varpi} \ln\frac{\lambda_t^2f^2\sin2\vartheta}{2\Lambda^2}}{2\sin^2\vartheta+\frac{\varpi}{1+\varpi} \ln\cos^2\vartheta}=\frac{f^2}{v^2}-\frac{3\lambda_t^4f^2}{4\pi^2m_h^2}\ln\frac{f}{v}\ldots \\
\Delta_{\Lambda^2}=&\,\frac{\frac{\varpi/2}{1+\varpi}}{\sin^2\vartheta+\frac{\varpi}{1+\varpi} \ln\cos^2\vartheta}=\frac{3\lambda_t^4}{4\pi^2}\frac{f^2}{m_h^2}+\ldots~\\
\Delta_{f^2}=&\,\frac{1+\varpi\ln\left(\cos\vartheta\sin\vartheta \right)}{2\sin^2\vartheta+2\varpi\left(\sin^2\vartheta+\ln\cos\vartheta \right)}=\frac{f^2}{v^2}-\frac{3\lambda_t^4f^2}{4\pi^2m_h^2}\ln\frac{f}{v}\ldots~.\label{eq:THf2TuneExp}
\end{align}
We see that as $f/v\sim f/m_h$ that all of the tunings are of the same size, $f^2/v^2$. The tuning due to the cutoff is not sensitive to the tree-level tuning, but only the loop effects and so it is reduced by a loop factor relative to the others. But in every case the size of the loop factor contribution is similar. 

\subsection{Tuning in the SU(4$)_c$ Color Model\label{ss.TuningSU4}}
What changes in the SU(4$)_c$ model? We assume that the scale of SU(4$)_c$ color breaking $w$ is well above the scale $f$ that breaks the approximate SU(4) global symmetry of the Higgs potential. This means that the tree-level Higgs potential is essentially unchanged from the standard Twin Higgs analysis except that $m^2=\delta_{H\Phi}w^2$.

The significant change comes from the one-loop contributions to the potential. This has the form given in Eq.~\eqref{e.CWpot} with
\begin{align}
m_{t_A}^2=\lambda_t^2f^2\sin^2\frac{\sqrt{\bm{h}^\dag\bm{h}}}{f}~, \ \ \ m_{t_B}^2=\lambda_t^2f^2\cos^2\frac{\sqrt{\bm{h}^\dag\bm{h}}}{f}~.
\end{align}
These come with three and four colors, respectively. The remaining, colorless, fermions have masses given by
\begin{align}
2m_{T_\pm}^2=&\, w^2\left(\lambda_{\widehat{U}}^2+\lambda_{\widehat{Q}}^2\right)+\lambda_t^2f^2\sin^2\frac{\sqrt{\bm{h}^\dag\bm{h}}}{f}\nonumber\\
&\pm\sqrt{\left[w^2\left(\lambda_{\widehat{U}}^2+\lambda_{\widehat{Q}}^2\right)+\lambda_t^2f^2\sin^2\frac{\sqrt{\bm{h}^\dag\bm{h}}}{f} \right]^2-4w^4\lambda_{\widehat{U}}^2\lambda_{\widehat{Q}}^2}~.
\end{align}
Note that these masses satisfy
\begin{align}
m_{T_+}m_{T_-}=&\,w^2\lambda_{\widehat{U}}\lambda_{\widehat{Q}}\\
m_{T_+}^2+m_{T_-}^2=&\,w^2\left(\lambda_{\widehat{U}}^2+\lambda_{\widehat{Q}}^2\right)+\lambda_t^2f^2\sin^2\frac{\sqrt{\bm{h}^\dag\bm{h}}}{f}\\
m_{T_+}^2-m_{T_-}^2=&\,\sqrt{\left[w^2\left(\lambda_{\widehat{U}}^2+\lambda_{\widehat{Q}}^2\right)+\lambda_t^2f^2\sin^2\frac{\sqrt{\bm{h}^\dag\bm{h}}}{f} \right]^2-4w^4\lambda_{\widehat{U}}^2\lambda_{\widehat{Q}}^2}~.
\end{align}
We see that in order for the masses to be equal we must have 
\beq
w^2\left(\lambda_{\widehat{U}}\pm\lambda_{\widehat{Q}} \right)^2+m_{t_A}^2=0~,
\eeq
which is impossible. 
Assuming that $\lambda_{\widehat{U}}>\lambda_{\widehat{Q}}$ (identical conditions follow from the opposite assumption) we find that
\begin{align}
    2w^2\lambda_{\widehat{U}}^2=&\,m_{T_+}^2+m_{T_-}^2-m_{t_A}^2+\sqrt{\left(m_{T_+}^2+m_{T_-}^2-m_{t_A}^2\right)^2-4m_{T_+}^2m_{T_-}^2}~,\\
    2w^2\lambda_{\widehat{Q}}^2=&\,m_{T_+}^2+m_{T_-}^2-m_{t_A}^2-\sqrt{\left(m_{T_+}^2+m_{T_-}^2-m_{t_A}^2\right)^2-4m_{T_+}^2m_{T_-}^2}~.
\end{align}
From this we see that the limit of $\lambda_{\widehat{U}}=\lambda_{\widehat{Q}}$ implies that 
\beq
\left(m_{T_+}\pm m_{T_-} \right)^2=m_{t_A}^2~.
\eeq
Of course, experimental considerations forbid the sum of the masses to be this small. 

To use the methods outlined above, we treat these masses as function of $x=\sqrt{\bm{h}^\dag\bm{h}}$. We find
\begin{align}
    \frac{dm_{T_\pm}^2}{dx}=&\pm\lambda_t^2f\frac{m_{T_\pm}^2}{m_{T_+}^2-m_{T_-}^2}\sin\frac{2x}{f}\\
    \frac{d^2m_{T_\pm}^2}{dx^2}=&\pm2\lambda_t^2\frac{m_{T_\pm}^2}{m_{T_+}^2-m_{T_-}^2}\left[\cos\frac{2x}{f}-\lambda_t^2f^2\sin^2\frac{2x}{f}\frac{m_{T_\mp}^2}{(m_{T_+}^2-m_{T_-}^2)^2} \right]
\end{align}

This leads to
\begin{align}
V_\text{CW}'(x)=&\,\frac{\lambda_t^2f}{8\pi^2}\sin\frac{2x}{f}\left( 4\lambda_t^2f^2\cos^2\frac{x}{f}\ln\frac{\lambda_t^2f^2\cos^2\frac{x}{f}}{\Lambda^2}-3\lambda_t^2f^2\sin^2\frac{x}{f}\ln\frac{\lambda_t^2f^2\sin^2\frac{x}{f}}{\Lambda^2}\right.\nonumber\\
&\left.-\frac{m_{T_+}^4}{m_{T_+}^2-m_{T_-}^2}\ln\frac{m_{T_+}^2}{\Lambda^2}+ \frac{m_{T_-}^4}{m_{T_+}^2-m_{T_-}^2}\ln\frac{m_{T_-}^2}{\Lambda^2}\right)~,
\end{align}
and 
\begin{align}
    V_\text{CW}''(x)=&-\frac{\lambda_t^2}{8\pi^2}\left\{ \lambda_t^2f^2\left[\sin^2\frac{2x}{f}\left(7+\frac{m_{T_+}^4+m_{T_-}^4}{(m_{T_+}^2-m_{T_-}^2)^2} \right) \right.\right.\\
    &-4\left(\cos\frac{2x}{f}+\cos\frac{4x}{f}\right)\ln\frac{\lambda_t^2f^2\cos^2\frac{x}{f}}{\Lambda^2} +3\left.\left(\cos\frac{2x}{f}-\cos\frac{4x}{f}\right)\ln\frac{\lambda_t^2f^2\sin^2\frac{x}{f}}{\Lambda^2}\right]\nonumber\\
    &+\frac{m_{T_+}^4}{(m_{T_+}^2-m_{T_-}^2)^2}\left[2(m_{T_+}^2-m_{T_-}^2)\cos\frac{2x}{f}+\lambda_t^2f^2\sin^2\frac{2x}{f}\frac{m_{T_+}^2-3m_{T_-}^2}{m_{T_+}^2-m_{T_-}^2} \right]\ln\frac{m_{T_+}^2}{\Lambda^2}\nonumber\\
    &-\left.\frac{m_{T_-}^4}{(m_{T_+}^2-m_{T_-}^2)^2}\left[2(m_{T_+}^2-m_{T_-}^2)\cos\frac{2x}{f}+\lambda_t^2f^2\sin^2\frac{2x}{f}\frac{m_{T_-}^2-3m_{T_+}^2}{m_{T_+}^2-m_{T_-}^2} \right]\ln\frac{m_{T_-}^2}{\Lambda^2}\right\}\nonumber
\end{align}
Using these results we find that the vacuum constraint for this scenario is
\begin{align}
w^2\delta_{H\Phi}=&\,f^2\delta_H\cos2\vartheta-\frac{\lambda_t^2}{16\pi^2}\left(4\lambda_t^2f^2\cos^2\vartheta\ln\frac{\lambda_t^2f^2\cos^2\vartheta}{\Lambda^2}-3\lambda_t^2f^2\sin^2\vartheta\ln\frac{\lambda_t^2f^2\sin^2\vartheta}{\Lambda^2} \right.\nonumber\\
&\left.-\frac{m_{T_+}^4}{m_{T_+}^2-m_{T_-}^2}\ln\frac{m_{T_+}^2}{\Lambda^2}+\frac{m_{T_-}^4}{m_{T_+}^2-m_{T_-}^2}\ln\frac{m_{T_-}^2}{\Lambda^2}\right)\label{e.su4Vac}
\end{align}

The $\mu^2$ terms is 
\begin{align}
  \mu^2=&\,2\left(\delta_Hf^2-\delta_{H\Phi}w^2 \right)\nonumber\\
  &-\frac{\lambda_t^2}{8\pi^2}\left[4\lambda_t^2f^2\ln\frac{\lambda_t^2f^2}{\Lambda^2}-\frac{w^2}{\lambda_{\widehat{U}}^2-\lambda_{\widehat{Q}}^2}\left( \lambda_{\widehat{U}}^4\ln\frac{w^2\lambda_{\widehat{U}}^2}{\Lambda^2}- \lambda_{\widehat{Q}}^4\ln\frac{w^2\lambda_{\widehat{Q}}^2}{\Lambda^2}\right)\right]~.
\end{align}
Note that this has a finite limit for $\lambda_{\widehat{U}}=\lambda_{\widehat{Q}}$
\beq
\lim_{\lambda_{\widehat{U}}\to\lambda_{\widehat{Q}}}\mu^2=2\left(\delta_Hf^2-\delta_{H\Phi}w^2 \right)-\frac{\lambda_t^2}{8\pi^2}\left[4\lambda_t^2f^2\ln\frac{\lambda_t^2f^2}{\Lambda^2}-w^2\lambda_{\widehat{Q}}^2\left(1+4\ln\frac{w\lambda_{\widehat{Q}}}{\Lambda} \right) \right]~.
\eeq
The physical Higgs mass, after using the vacuum constraint, is found to be
\begin{align}
    m_h^2=&\,2f^2\sin^22\vartheta\left\{\delta_H-\frac{\lambda_t^4}{32\pi^2}\left[8+\frac{2m_{T_+}^2m_{T_-}^2}{(m_{T_+}^2-m_{T_-}^2)^2} +7\ln\frac{\lambda_t^2f^2}{\Lambda^2} +\ln\left( \cos^8\vartheta\sin^6\vartheta\right)\right.\right.\nonumber\\
  &\left.\left.+\ln\frac{m_{T_+}m_{T_-}}{\Lambda^2}  +\frac{(m_{T_+}^2+m_{T_-}^2)\left[\left(m_{T_+}^2-m_{T_-}^2 \right)^2-2m_{T_+}^2m_{T_-}^2\right]}{\left(m_{T_+}^2-m_{T_-}^2\right)^3}\ln\frac{m_{T_+}}{m_{T_-}}\right] \right\}~.
\end{align}

These results allow us to calculate the tuning in this model. We first evaluate the derivatives and then apply the constraints. It is useful to record the vacuum constraint in terms of the Higgs mass
\begin{align}
w^2\delta_{H\Phi}=&\frac{m_h^2\cos2\vartheta}{2\sin^22\vartheta}+\frac{\lambda_t^2}{32\pi^2}\left\{2\lambda_t^2f^2\cos2\vartheta\left(4+\frac{m_{T_+}^2m_{T_-}^2}{(m_{T_+}^2-m_{T_-}^2)^2} \right)+f^2\lambda_t^2\ln\frac{m_{T_+}m_{T_-}\sin^6\vartheta}{f^2\lambda_t^2\cos^8\vartheta} \right.\nonumber\\
&+2w^2\left(\lambda_{\widehat{U}}^2+\lambda_{\widehat{Q}}^2 \right)\ln\frac{m_{T_+}m_{T_-}}{\Lambda^2}-\frac{1}{m_{T_+}^2-m_{T_-}^2}\left[2\left(m_{T_+}^4+m_{T_-}^4 \right)\right.\nonumber\\
&\left.\left.+\lambda_t^2f^2\cos2\vartheta\left(1-\frac{2m_{T_+}^2m_{T_-}^2}{(m_{T_+}^2-m_{T_-}^2)^2} \right)\left(m_{T_+}^2+m_{T_-}^2\right) \right]\ln\frac{m_{T_-}}{m_{T_+}}\right\}~.
\end{align}
To understand the tuning results, it is useful to determine the leading order behavior:
\begin{align}
w^2\delta_{H\Phi}\approx&\, \frac{m_h^2 f^2}{4v^2}+\frac{\lambda_t^2w^2}{16\pi^2}\left[ \left(\lambda_{\widehat{U}}^2+\lambda_{\widehat{Q}}^2\right)\ln\frac{w^2\lambda_{\widehat{U}}\lambda_{\widehat{Q}}}{\Lambda^2}+\frac{\lambda_{\widehat{U}}^4+\lambda_{\widehat{Q}}^4}{\lambda_{\widehat{U}}^2-\lambda_{\widehat{Q}}^2}\ln\frac{\lambda_{\widehat{U}}}{\lambda_{\widehat{Q}}} \right]
\end{align}
We also express $\mu^2$ with the vacuum constraint and the formula for the Higgs mass applied as
\begin{align}
&\mu^2=\frac{m_h^2}{2\cos^2\vartheta}+\frac{\lambda_t^2}{8\pi^2}\left\{8\lambda_t^2f^2\left(\sin^2\vartheta+\ln\cos\vartheta \right)+2\lambda_t^2f^2\sin^2\vartheta\frac{m_{T_+}^2m_{T_-}^2}{(m_{T_+}^2-m_{T_-}^2)^2} \right.\nonumber\\
&+\frac{1}{m_{T_+}^2-m_{T_-}^2}\left[m_{T_+}^4+m_{T_-}^4-\lambda_t^2f^2\sin^2\vartheta(m_{T_+}^2+m_{T_-}^2)\left(1 - \frac{2m_{T_+}^2m_{T_-}^2}{(m_{T_+}^2-m_{T_-}^2)^2} \right)\right]\ln\frac{m_{T_-}}{m_{T_+}}\nonumber\\
&\left.\hspace{1cm}-w^2\frac{\lambda_{\widehat{U}}^4+\lambda_{\widehat{Q}}^4}{\lambda_{\widehat{U}}^2-\lambda_{\widehat{Q}}^2}\ln\frac{\lambda_{\widehat{Q}}}{\lambda_{\widehat{U}}}\right\}~.
\end{align}
Note that this recovers $m_h^2=2\mu^2$ in the $\vartheta\to0$ limit. Indeed, as this form of $\mu^2$ is to be used in the denominator of the tuning calculations only, we expand it in $v/f\ll1$ to find 
\begin{align}
    \mu^2=\frac{m_h^2}{2}+\frac{3\lambda_t^4v^2}{16\pi^2}+\mathcal{O}\left(m_h^2\frac{v^2}{f^2} \right)
\end{align}

The equations are quite long and so we only present the leading terms in $v/f$. We find
\begin{align}
\Delta_{\delta_H}=&\,\frac{2\delta_Hf^2}{\mu^2}\\
\Delta_{f^2}=&\,\Delta_{\delta_H}-\frac{\lambda_t^4f^2}{2\pi^2\mu^2}\left(1+\ln\frac{\lambda_t^2f^2}{\Lambda^2} \right)\\
\Delta_{\delta_{H\Phi}}=&-\frac{2w^2\delta_{H\Phi}}{\mu^2}\\
\Delta_{w^2}=&\,\Delta_{\delta_{H\Phi}}\nonumber\\
&+\frac{w^2\lambda_t^2}{8\pi^2\mu^2}\left[\left(\lambda_{\widehat{U}}^2+\lambda_{\widehat{Q}}^2 \right)\left(1+\ln\frac{w^2}{\Lambda^2} \right)+\frac{1}{\lambda_{\widehat{U}}^2-\lambda_{\widehat{Q}}^2}\left(\lambda_{\widehat{U}}^4\ln\lambda_{\widehat{U}}^2-\lambda_{\widehat{Q}}^4 \ln\lambda_{\widehat{Q}}^2\right) \right]\\
    \Delta_{\Lambda^2}=&\,\frac{\lambda_t^2}{8\pi^2\mu^2}\left[4\lambda_t^2f^2-w^2\left(\lambda_{\widehat{U}}^2+\lambda_{\widehat{Q}}^2 \right) \right]\\
\Delta_{\lambda_{\widehat{U}}}=&\,\frac{\lambda_t^2w^2\lambda_{\widehat{U}}^2}{4\pi^2\mu^2} \left[\frac{\lambda_{\widehat{U}}^2}{\lambda_{\widehat{U}}^2-\lambda_{\widehat{Q}}^2}+\ln\frac{w^2\lambda_{\widehat{U}}\lambda_{\widehat{Q}}}{\Lambda^2}+\left(1-\frac{2\lambda_{\widehat{Q}}^4}{(\lambda_{\widehat{U}}^2-\lambda_{\widehat{Q}}^2)^2} \right)\ln\frac{\lambda_{\widehat{U}}}{\lambda_{\widehat{Q}}} \right]\\
\Delta_{\lambda_{\widehat{Q}}}=&\,\frac{\lambda_t^2w^2\lambda_{\widehat{Q}}^2}{4\pi^2\mu^2} \left[-\frac{\lambda_{\widehat{Q}}^2}{\lambda_{\widehat{U}}^2-\lambda_{\widehat{Q}}^2}+\ln\frac{w^2\lambda_{\widehat{U}}\lambda_{\widehat{Q}}}{\Lambda^2}-\left(1-\frac{2\lambda_{\widehat{U}}^4}{(\lambda_{\widehat{U}}^2-\lambda_{\widehat{Q}}^2)^2} \right)\ln\frac{\lambda_{\widehat{U}}}{\lambda_{\widehat{Q}}} \right]
\end{align}
In these expressions the $f/v$ tuning of tree-level potential persists. We also see that for the cutoff and $w^2$ that there is a large loop-level contribution that goes like $w^2/m_h^2$ divided by 4$\pi^2$. If the new Yukawas are close to order one then this can be a significant contribution to the tuning.

\bibliographystyle{JHEP}
\bibliography{BIB}{}

\end{document}